\documentclass[fleqn]{2023SCGE}
\usepackage{graphicx,subfigure,amsmath,amsfonts,amssymb,multirow,mathrsfs,hyperref}
\usepackage{ulem,soul}
\begin{document}
\ensubject{subject}

\ArticleType{Article}
\SpecialTopic{SPECIAL TOPIC: }
\Year{2026}
\Month{November}
\Vol{00}
\No{0}
\DOI{00}
\ArtNo{000000}

\title{Fast Radio Bursts Trace Cosmic Star Formation with Little Delay}

\author[1]{Yi-Ying Wang}{}
\author[1]{Yin-Jie Li}{}
\author[1,2]{Yi-Zhong Fan}{yzfan@pmo.ac.cn}

\AuthorMark{Wang Y-Y}

\AuthorCitation{Wang Y-Y, Li Y-J, Fan Y-Z}

\address[1]{Purple Mountain Observatory, Chinese Academy of Sciences, Nanjing 210023, People's Republic of China}
\address[2]{School of Astronomy and Space Science, University of Science and Technology of China, Hefei, Anhui 230026, People's Republic of China}

\abstract{The progenitor channels of fast radio bursts (FRBs) remain debated, with a central question being whether their cosmic rate traces star formation promptly or instead follows it with the long time-delay characteristic of compact-binary mergers. We perform a forward-modeling, hierarchical Bayesian analysis of the CHIME/FRB population, jointly fitting the catalog sample, baseband fluences, and localized host redshifts, while self-consistently incorporating the survey selection function through the injection framework. 
Adopting the classical cosmic star formation history (SFH) with a range of delay-time models, we find that the reconstructed FRB rate robustly tracks the SFH peak. All models exclude characteristic delays $\Delta t_d > 1.0$ Gyr at the 90\% credible level, with inferred median peak offsets in the range $\Delta t_d = 0.1$--$0.3$ Gyr that remain consistent with a prompt, zero-delay origin at the 2$\sigma$ level.
This finding disfavors the previously reported multi-Gyr delays, strongly indicating that the central engine is linked to young remnants (likely magnetars) from stellar collapse and, potentially, to the prompt merger of compact objects.}



\maketitle
\begin{multicols}{2}
\section{Introduction.} \label{sec:intro}
Fast radio bursts (FRBs) are millisecond-duration radio transients of extragalactic and cosmological origin. Since their discovery \cite{2007Sci...318..777L}, more than 4000 FRB sources have been reported\footnote{\url{https://blinkverse.zero2x.org}}. A fraction of FRBs repeat, whereas the majority have so far been detected only as apparent one-off events. Nevertheless, the nature of their progenitors remains an open question \cite{2019ARA&ACordes, 2019PhRPlatts, 2021XiaoD, 2022A&ARvPetroff, 2023ZhangB}. For instance, the detection of an FRB-like burst from the Galactic magnetar SGR 1935+2154 \cite{2020NaturCHIME, 2020NaturBochenek, 2021NatAsKirsten} established magnetars as a viable engine. In contrast, the repeating source FRB 20200120E \cite{2021Bhardwaj, 2022NaturKirsten} localized to a globular cluster of the nearby galaxy M81 indicates that some FRBs can be associated with old stellar populations. Proposed formation channels can accordingly be grouped into two broad pathways with distinct cosmic histories: (i) a prompt channel traces the cosmic star formation history (SFH) with negligible delay \cite{2020MNRASHashimoto}, such as the magnetars formed in core-collapse supernovae \cite{2010ApJ...719L.204W,LiShang2026PRL} or supramassive neutron stars collapsing into black holes \cite{2014A&AFalcke}; and (ii) a delayed channel driven by compact-binary mergers that convolves star formation rate (SFR) with a broad delay-time distribution (DTD) \cite{2020NaturZhangB}, such as binary neutron-star mergers \cite{2013PASJTotani,2016ApJLWangJS} or magnetars produced in mergers or in the accretion-induced collapse of white dwarfs \cite{1992ApJDuncan,1992Natur.357..472U,2021ApJKremer}. Because these channels predict distinct cosmic rate histories, the redshift evolution of the FRB population provides a powerful probe of the progenitor age and formation channels.

Motivated by this connection, a number of population studies have modeled the FRB redshift distribution and reported a preference for a long characteristic delay. The derived delay times are broadly consistent with the DTD of short gamma-ray bursts (sGRBs), thus favoring an old, merger-like population \cite{2022JCAPQiangDC, 2022ApJZhangRachel, 2024ApJChenJH, 2024ApJLinHN, 2025ApJChampati, 2025A&AZhangKJ, 2025ApJZhouH, 2026arXiv26SuZW}. If robust, such a conclusion would place FRBs and compact-binary mergers on a common evolutionary footing. However, these inferences are sensitive to the treatment of selection effects, including the use of Lynden-Bell's $c^{-}$ method \cite{1971MNRASLynden-BellD} or the gray-zone fluence criterion. Indeed, forward-modeling analyses that explicitly incorporate the instrumental response tend to recover populations consistent with the SFH \cite{2022MNRASJames, 2023ApJShin}, leaving open the possibility that the apparent preference for large delays is driven by modeling choices rather than by the data themselves.

The Canadian Hydrogen Intensity Mapping Experiment (CHIME) has transformed this problem by providing the first large FRB sample accompanied by a mock population of synthetic FRBs \cite{2021ApJSCHIME, 2023AJMerryfield}. In addition to the catalog-level sample, CHIME baseband events provide improved burst characterization \cite{2024ApJCHIME}, while the subset of localized FRBs with host redshifts anchors the dispersion measure (DM) -- redshift relation. In this work, we perform a forward-modeling and unbinned hierarchical population analysis of these joint samples to reconstruct the cosmic evolution of the FRB rate, and thereby discriminate among candidate formation channels. 

\section{Methods} \label{sec:method}
Our goal is to reconstruct the cosmic FRB volumetric rate $\mathcal{R}(z)$
directly from survey data. We forward model a proposed population through the full instrument response and evaluate it against the detections. Given a set of observations ${d_i}$ from $N_{\rm det}$ FRB sources, the likelihood of the observed data for a population described by the hyperparameters $\bf \Lambda$ follows the inhomogeneous Poisson process \cite{2019MNRASMandel, 2019PASAThrane}
\begin{equation}\label{eq:hier}
\mathcal{L}(\{d\}\,|\,{\bf \Lambda})\propto
N_{\rm exp}^{N_{\rm det}}\,e^{-N_{\rm exp}}
\prod_{i=1}^{N_{\rm det}}\frac{1}{\xi({\bf \Lambda})}
\int \mathcal{L}(d_i|\theta)\,\pi(\theta|{\bf \Lambda})\,{\rm d}\theta,
\end{equation}
where $N_{\rm exp}=N\xi({\bf \Lambda})$ is the expected number of detections out of $N$ bursts occurring during the survey, $\theta=\{z,F,\mathrm{DM}\}$ are the latent parameters of an individual burst, $\pi(\theta|{\bf \Lambda})=\pi(F|z,\Lambda_1)\, \pi(\mathrm{DM}|z,\Lambda_2)\,\pi(z|\Lambda_3)$ is the normalized population probability density with hyperparameters ${\bf \Lambda}=\{\Lambda_1,\Lambda_2,\Lambda_3\}$, $\Lambda_1=\{ \gamma, E_{\rm char}, \alpha\}$ governs the fluence and energy distributions, $\Lambda_2=\{\mu_{\rm host}$ , $\sigma_{\rm host}\}$ represents the host-galaxy DM model, and $\Lambda_3$ denotes the parameters of each cosmic rate-evolution model. 

A key ingredient is the detection efficiency 
\begin{equation}
\xi({\bf \Lambda})=\int P_{\rm det}(\theta)\,\pi(\theta|{\bf \Lambda})\, {\rm d}\theta.
\end{equation}
Instead of adopting an analytic threshold model, we estimate it with the CHIME/FRB injection system \cite{2023AJMerryfield}\footnote{\url{https://chime-frb-open-data.github.io/injections/}}, where synthetic bursts drawn from a reference distribution $P_{\rm inj}(\theta)$ were injected into the live intensity data stream and processed by the same pipeline as real events, so that
\begin{equation}\label{eq:xi}
\hat{\xi}({\bf \Lambda})=\frac{1}{N_{\rm inj}}
\sum_{j\,\in\,{\rm found}}\frac{\pi(\theta_j|{\bf \Lambda})}{P_{\rm inj}(\theta_j)}.
\end{equation}
Since $\xi({\bf\Lambda})$ naturally propagates the beam response, radio-frequency-interference excision, and pipeline completeness into the likelihood, it effectively quantifies the survey selection effects.

A second key ingredient is the joint use of three data types within a single likelihood. The majority of bursts come from the CHIME/FRB catalog~1 \cite{2021ApJSCHIME} (Type~1) for which DM is measured but the fluence is known only as a lower limit, with no host redshift. A subset of bursts (Type~2) has precise fluences $F_{\rm obs}\pm\sigma_F$ measured from baseband data~\cite{2024ApJCHIME} but still lacks a host association, and hence a redshift. A further four FRBs (Type~3) have baseband information and secure host identifications with spectroscopic redshifts. Although small in number, these localized bursts anchor the DM--redshift relation and break degeneracies among the rate evolution, the energy distribution, and the $\rm DM_{host}$ distribution. The likelihood for a single event $\mathcal L(d|{\bf \Lambda})$ combines the information from $\rm DM$ and $F$. Because the sample is divided into three categories, the corresponding likelihoods of $F$ differ among them. For Type~1 sources, for which only a lower limit on $F$ is available, the model is evaluated directly in the observed $(\mathrm{S/N},\mathrm{DM})$ plane. The predicted detection density is evaluated through a weighted Gaussian kernel-density estimate $K(\mathrm{S/N},\mathrm{DM})$ over the recovered injections, with each injection weighted by 
\begin{equation}\label{eq:weights}
w_j \propto \pi(\theta_j|{\bf \Lambda})/P_{\rm inj}(\theta_j).
\end{equation}
The Gaussian kernels have covariance $h^2C$ in the $(\ln \mathrm{S/N},\, \ln \mathrm{DM})$ coordinate, where $C$ is the unweighted covariance matrix of the recovered injections and the fiducial bandwidth is $h=0.2$. The inferred posterior on $\bf \Lambda$ is stable for $h\in [0.05, 0.4]$. Normalizing the weights over the recovered injections makes $K$ a properly normalized probability density in the $(\mathrm{S/N},\mathrm{DM})$ plane, i.e. the normalization automatically divides out $\xi({\bf \Lambda})$.

For Type~2 and Type~3 sources, the fluence likelihoods $\mathcal{L}_{2,3}(F_i|F)$ follow Gaussian distributions, centered on $\mu=F_{\rm obs}$ with width $\sigma=\sigma_{F_{\rm obs}}$. The DM likelihoods $\mathcal{L}_{2,3}(\mathrm{DM}_i|\Lambda_2,z)$ are evaluated at the measured $\mathrm{DM}_i$ of each observed burst and interpolated linearly on the redshift grid $z$. Because the uncertainties in the DM contributions of the host galaxies and the diffuse cosmic baryons greatly exceed the measurement error $\Delta \rm DM_{obs}$, the latter is treated as a delta function. For Type~2 sources, this quantity is marginalized over the population redshift distribution $\pi(z|\Lambda_3)$. For Type~3 sources, it is evaluated at the spectroscopic redshift. The total likelihood is therefore given by
\begin{equation}
\begin{aligned}
&\mathcal{L}(\{d\}|{\bf \Lambda}, N)\propto N_{\rm exp}^{N_{\rm det}}e^{-N_{\rm exp}} \prod_{i}^{N_{\rm det,1}} K(\mathrm{S/N}_i,\mathrm{DM}_i|{\bf \Lambda}) \\
&\times \prod_{j}^{N_{\rm det,2}} \frac{1}{\xi({\bf \Lambda})}\int{\rm d}z\pi(z|\Lambda_3) \mathcal{L}_2({{\rm DM}_j|\Lambda_2}, z) \\ 
& \qquad \qquad \  \ \times\int {\rm d}F\pi(F|\Lambda_1,z)\mathcal{L}_{2}(F_j|F) \\
& \times  \prod_{k}^{N_{\rm det,3}} \frac{1}{\xi({\bf \Lambda})} \pi(z_k|\Lambda_3) \pi({\rm DM}_k|\Lambda_2,z_k) \\
& \qquad \qquad \  \ \times\int {\rm d}F\pi(F|\Lambda_1,z_k)\mathcal{L}_{3}(F_k|F),
\end{aligned}
\end{equation}

where the fluence integral extends from 0.2 to $10^{5.7} \, \rm Jy \,ms$, and the redshift integral extends from 0 to 5. 

To ensure the accuracy of our inference, we require $N_{\rm eff}>4 N_{\rm det}$ \cite{2019RNAASFarr},
where the effective sample size is
\begin{equation}
N_{\rm eff}({\bf \Lambda})=\frac{\big(\sum_{j\,\in\,{\rm found}}w_j\big)^{2}} {\sum_{j\,\in\,{\rm found}}w_j^{2}}.
\label{eq:neff}
\end{equation} Following the selection criteria of Ref. \cite{2023ApJShin}, the final sample comprises 225 sources. We provide Supplementary Materials to detail the parameter definitions, prior ranges, and the sample selections.

The population model is specified by the distributions of $F$, DM, and $z$. The isotropic equivalent energy is modeled with a Schechter function with slope $\gamma$ and cutoff $E_{\rm char}$ above a pivot $E_{\rm pivot}=10^{38}\,$erg, chosen conservatively because it lies below the previously inferred lower bound $E_{\rm min} > 10^{38.5} \, \rm erg $ (90\% credible level) \cite{2022MNRASJames}. The energy $E$ is converted to the observed fluence $F$ with spectral index $\alpha$. The DM contribution of the host galaxy is modeled with a log-normal distribution with parameters $(\mu_{\rm host},\sigma_{\rm host})$, while the Galactic and cosmic terms follow established models (see the Supplementary Materials for details). The central quantity of the analysis is the volumetric rate, obtained by convolving the cosmic star-formation history $\psi(z)$ \cite{2014ARA&A..52..415M} with a delay-time distribution (DTD) $P(t_d)$,
\begin{equation} \label{eq:pi(z)}
  \pi(z|\Lambda_3) \propto \frac{1}{1+z} \frac{{\rm d}V}{{\rm d}z}
  \frac{\mathcal{R}(z)}{\mathcal{R}(0)},
\end{equation}
with
\begin{equation}\label{eq:conv}
\mathcal{R}(z) = \int_{t_d^{\min}}^{t(z)}
\big[\mathrm{SFR}(z_f)\big]^{n}\,P(t_d)\,{\rm d}t_d,
\end{equation}
where $z_f$ is the redshift at the formation time $t(z)-t_d$ and the index $n$ accounts for a possible nonlinear scaling with the SFH. To ensure that our conclusions are not driven by a particular parametrization, we consider a series of DTD models \cite{2011ApJVirgili, 2015MNRASWanderman}: (i) no delay, $P(t_d)=\delta(t_d)$; (ii) fixed delay, $P(t_d)=\delta(t_d-t_{d,0})$ with $t_{d,0}\in[1\, \rm Myr,\,10 \,Gyr]$; (iii) power-law delay, $P(t_d)\propto t_d^{\beta}$ over $[10\,\mathrm{Myr},\,t_H]$, the canonical form for compact-binary inspirals; (iv) log-normal delay, with central delay $\mu_{t_d}$ and width $\sigma_{t_d}$; and (v) Madau-Extended form, $\mathcal{R}(z)\propto(1+z)^{\lambda}\{1+[(1+z)/(1+z_p)]^{\kappa}\}^{-1}$, which relaxes the fixed Madau–Dickinson shape and captures the systematic uncertainty in the assumed SFH.

\begin{figure}[H]
\centering
\includegraphics[scale=0.45]{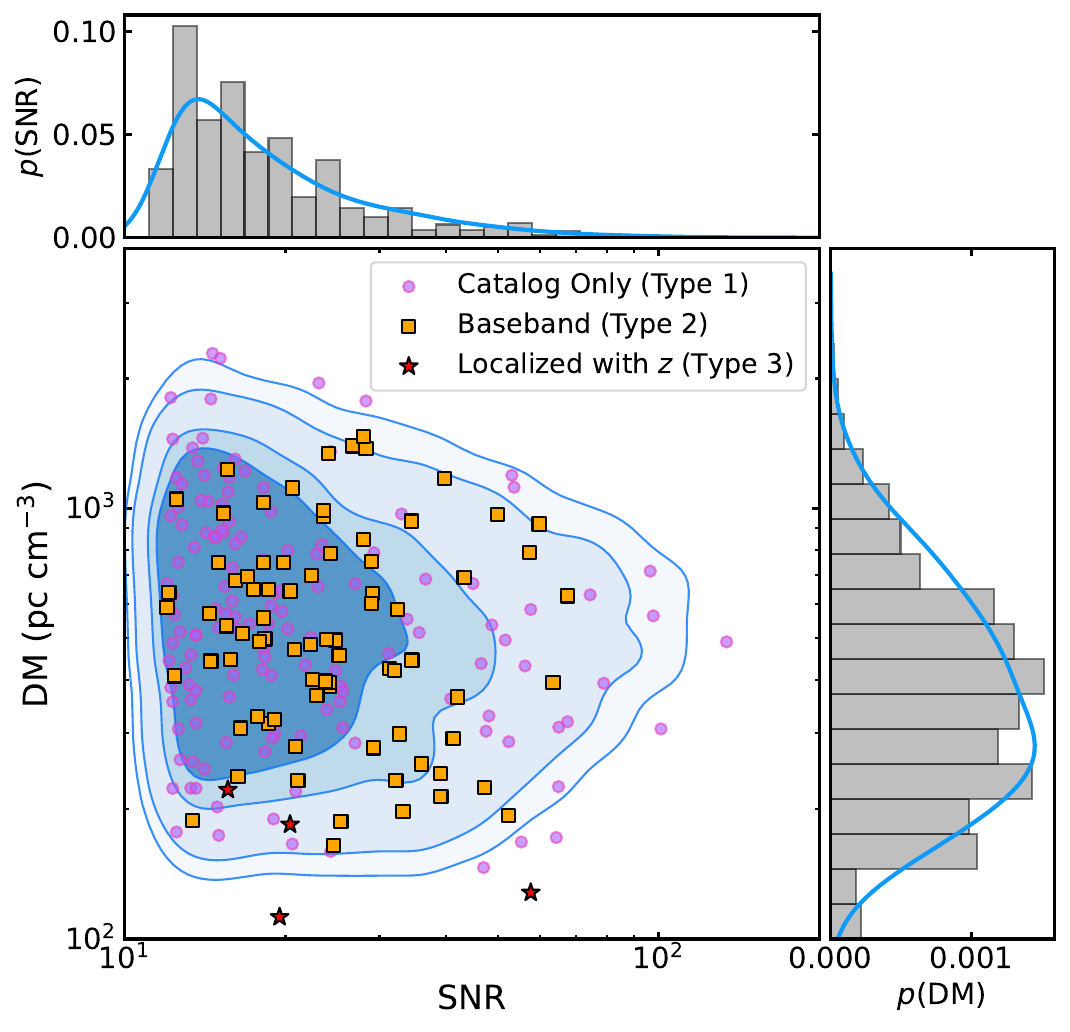}
\caption{Joint and marginal distributions of the S/N and DM for the CHIME/FRB Catalog~1 sample. The observed bursts are categorized into three subsamples. The blue contours and solid lines show the 2D and 1D probability density functions of the SFH-tracking model (evaluated at the best-fit parameters), which naturally incorporates the instrumental selection effects. The contours denote the 50\%, 68\%, 90\% and 95\% credible regions. For the DM distributions of Type~3 against the Type~1+2, $p_{\rm K\mbox{-}S}<0.01$ suggests the difference, which is expected rather than problematic. Our hierarchical likelihood does not assume the three subsamples share the same detected DM distribution but does assume that they share the underlying population parameters. The ($F$, DM) plane results are shown in the Supplementary Materials.}\label{fig:GoF}  
\end{figure}

\section{Results} \label{sec:result}
\cref{fig:GoF} demonstrates that the population model can reproduce the observed data. The model detection density (blue contours), obtained by propagating the inferred population through the CHIME/FRB selection effect, accurately recovers the joint distribution of the catalog sample in the ($\mathrm{S/N},\mathrm{DM}$) plane, including the marginal $\mathrm{S/N}$ and $\mathrm{DM}$ histograms (top and right panels). Based on the posterior predictive check (PPC) with Kolmogorov–Smirnov (K-S) statistics, the DM and S/N marginal distributions give $p_{\rm PPC} = 0.59$ and $0.17$, and the 2D distribution gives $p_{\rm PPC} = 0.17$, indicating no statistically significant discrepancy. The Type~2 bursts carry additional fluence information and are statistically consistent with the Type~1 bursts, with $p_{\rm K\mbox{-}S}=0.81$ and $p_{\rm A\mbox{-}D}=0.67$ for the K-S and Anderson–Darling (A-D) tests between the Type~1 and Type~2 DM distributions. Type~3 supply redshift information and share the underlying population parameters.

\begin{figure*}
\centering
\includegraphics[scale=0.45]{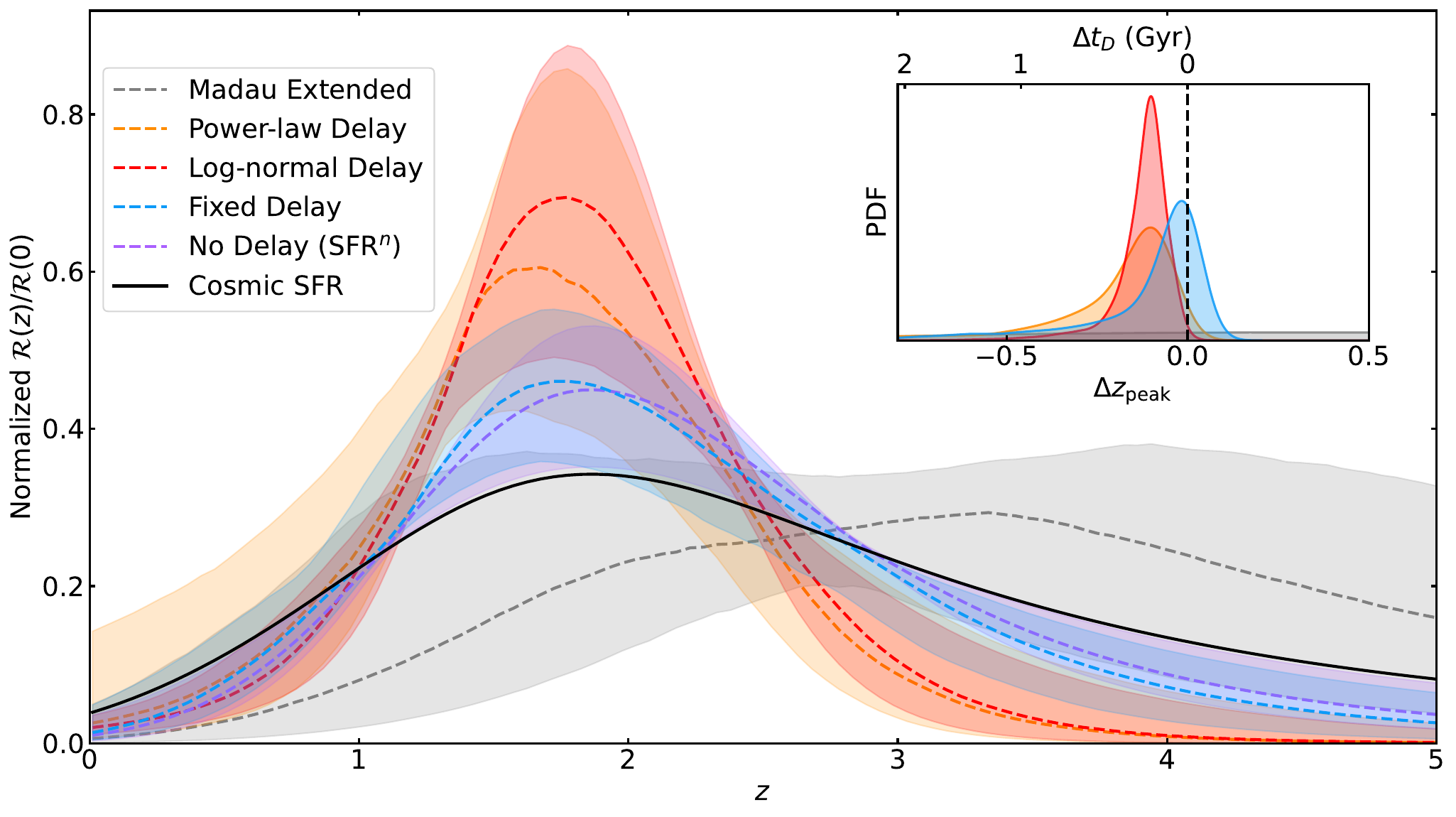}
\caption{Normalized intrinsic FRB event rate reconstructed for different DTD models. The shaded regions represent the 68\% credible intervals, while the dashed lines denote the medians. The inset shows the offset in redshift between the FRB rate peak and the cosmic SFR peak.}
\label{fig:rate} 
\end{figure*}

\begin{figure}[H]
\centering
\includegraphics[scale=0.45]{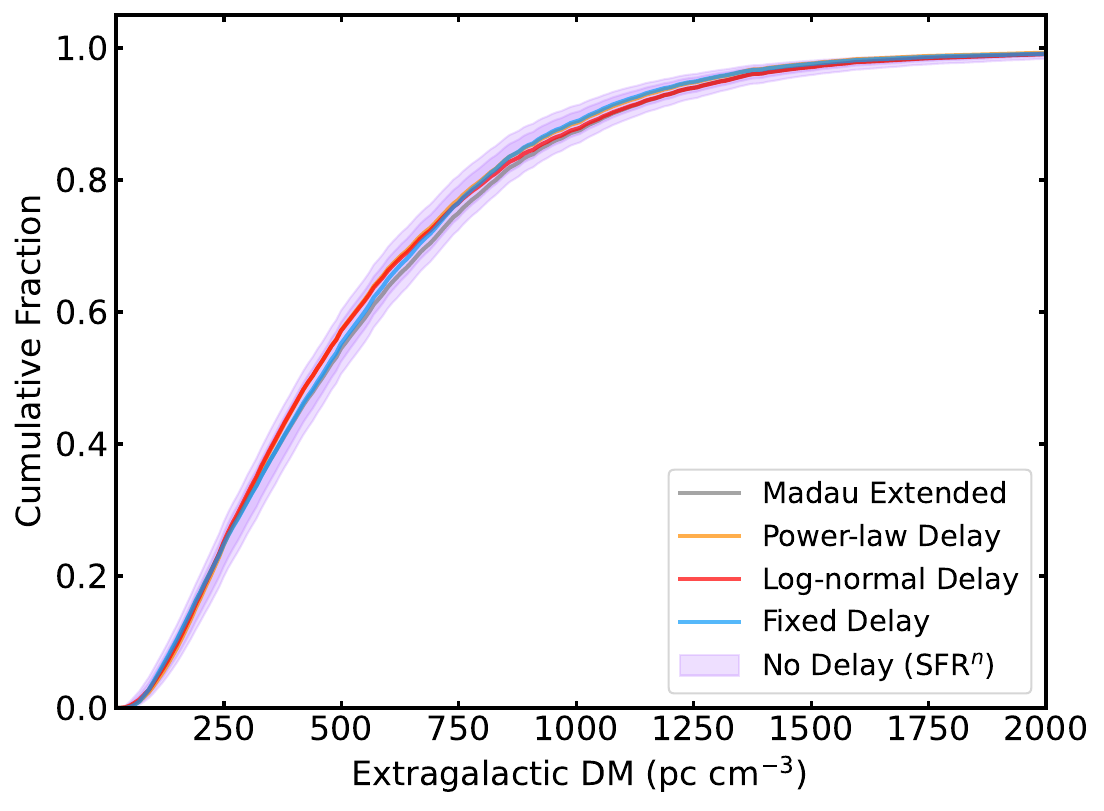}
\caption{The cumulative extragalactic DM for the different distribution models. The shaded regions represent the 68\% credible intervals of the SFH-tracking model. The solid lines represent the best-fit values.}
\label{fig:cum} 
\end{figure}

The central result is shown in \cref{fig:rate}, which compares the inferred normalized volumetric rate $\mathcal{R}(z)/\mathcal{R}(0)$ for all DTD models with the adopted cosmic SFH. Remarkably, the three physically motivated DTD models predict nearly identical rate evolutions, with peaks that closely track the assumed SFH peak. This coincidence reflects two factors. First, these models are anchored to the same fiducial SFH and share the same energy--fluence--redshift mapping. Second, the data favor short characteristic delays, such that convolving any of the DTDs with the SFH yields an event rate tightly coupled to the SFH shape.  
We quantify this agreement using summary statistics reported in two equivalent forms. These are the peak offset relative to the SFH, $\Delta z_{\rm peak}$, and the corresponding characteristic delay $\Delta t_d$, expressed as a lookback-time offset (top axis of the inset). Because this offset is measured directly from the inferred $\mathcal{R}(z)$, it is independent of the DTD parametrization assumed in each model. We find peak redshifts of $z_{\rm peak}=1.83^{+0.05}_{-0.20}$ (fixed), $1.77\pm0.05$ (log-normal), and $1.72^{+0.05}_{-0.25}$ (power-law), corresponding to lookback-time offsets relative to the SFH peak of $\Delta t_d=0.09^{+0.41}_{-0.09}$, $0.19\pm0.10$, and $0.29^{+0.57}_{-0.10}~\mathrm{Gyr}$, respectively. All three models under the assumed SFH exclude $\Delta t_d > 1.0$ Gyr at the 90\% credible levels. As shown in \cref{fig:cum}, the corresponding cumulative distributions of extragalactic DM are likewise mutually consistent and all reproduce the observed DM distribution within the credible interval of the no-delay model. These results serve as a consistency check, confirming that the different rate models remain compatible with the observed DM population.

\begin{table}[H]
\centering
\footnotesize
\caption{Secant Slopes Comparison}
\label{tab:slopes}
\tabcolsep 13pt
\begin{tabular}{lcc}
\toprule
Models &Slopes($z=1$) &Slopes($z=1.8$)\\ \hline
Power-law Delay &$0.23^{+0.05}_{-0.09}$ &$0.30\pm0.17$\\
Log-normal Delay &$0.21^{+0.05}_{-0.06}$ &$0.37^{+0.11}_{-0.13}$\\
Fixed Delay &$0.20^{+0.01}_{-0.03}$ &$0.22^{+0.06}_{-0.08}$\\
SFH tracking (${\rm SFR}^n$) &$0.20^{+0.01}_{-0.03}$ &$0.23^{+0.05}_{-0.08}$\\ 
Madau extended &$0.08^{+0.12}_{-0.05}$ &$0.10^{+0.09}_{-0.05}$\\
\bottomrule
\end{tabular}
\end{table}

The flexible ``Madau-Extended'' SFH model (gray) yields a rate peak at substantially higher redshift, $z_{\rm peak} = 2.83^{+1.26}_{-1.16}$, with a broad credible band indicating that the peak location of the DTD models relies on the assumed SFH. Based on the best-fit result, we find that only $\sim0.4$ detections are expected beyond $z\sim1.8$. For a quantitative comparison, we define the secant slope as $S(z)=[\widetilde{\mathcal R}(z)-\widetilde{\mathcal R}(0)]/z$, where $\widetilde{\mathcal R}(z)$ is the normalized volumetric rate shown in \cref{fig:rate}. Table~\ref{tab:slopes} reports the posterior medians and $68\%$ credible intervals of $S(z)$ at $z=1$ and $1.8$. The positive slope up to $z\sim1$, found consistently across all DTD models, is the feature most directly informed by the observed population, whereas its magnitude toward $z\sim1.8$ is more model dependent. We therefore infer that the rise of $\mathcal{R}(z)$ up to $z\sim1$ is directly constrained by the data. For the Madau-Extended model, the preference for high-$z$ activity still excludes multi-Gyr characteristic delays, which reflects a more data-driven result.

For the power-law DTD, the inferred slope $\beta=-1.11^{+0.51}_{-0.38}$ is compatible with the value $\beta\!\simeq\!-1$ expected for compact-binary mergers \cite{2019MNRASBeniamini}. However, it does not imply a long characteristic delay. With the minimum delay fixed at $t_{d,\min}=10$~Myr, the delay distribution is dominated by short timescales and is convolved with the broad cosmic SFH. The resulting event rate therefore remains tightly coupled to the SFH, with its peak nearly unchanged. This demonstrates that the slope of a power-law DTD alone is not a reliable diagnostic of the progenitor channel. In the Supplementary Materials, we further explore several scenarios of the power-law model with different fixed/free parameters, and obtain consistent results showing that the inferred event rate continues to closely follow the SFH.

Beyond the rate evolution, the remaining population hyperparameters are stable across all delay-time prescriptions. For example, the energy distribution is consistently characterized by a faint-end slope of $\gamma\simeq-1.8$ and a characteristic cutoff energy of $\log_{10}(E_{\rm char}/\mathrm{erg})\simeq42.1$. We also test for redshift evolution of the energy function and a broken-power-law alternative to the fiducial Schechter form. We find no evolution and no meaningful preference between the two forms. The inferred delay remains consistent with our fiducial result.  

We emphasize that $E_{\rm char}$ and $\Phi_0$ are defined with respect to the energy convention for fluence, the assumed rest-frame bandwidth $\Delta\nu=1\,\mathrm{GHz}$, and the single spectral index $\alpha$. The inferred local volumetric normalization is $\Phi_0\simeq1.3\times10^5 \, \rm Gpc^{-3} \, yr^{-1}$ at $z=0$ above $E_{\rm pivot} = 10^{38} \, \rm erg$. Comparison with Ref. \cite{2025A&AMengM} gives a consistent event rate, obtained under the same convention and energy threshold. 
Restricting the analysis to one-off FRBs yields nearly identical parameter constraints and provides a consistent normalized event rate, with only a modest decrease in $\Phi_0$, as expected. The only parameter exhibiting appreciable model dependence is the SFH exponent $n$, which ranges from $n\simeq1.5$ (fixed) to $n\simeq4.3$ (power-law, log-normal). This behavior reflects the degeneracy between $n$ and the DTD parameters, both of which modulate the high-redshift shape of $\mathcal{R}(z)$ but not the location of its peak. Such mildly super-linear scaling ($n>1$) is physically plausible if FRB progenitor formation is enhanced in the lower-metallicity environments characteristic of high-redshift star formation, as proposed for magnetar and long GRB progenitors \cite{2017ApJNicholl, 2021MNRASAloy, 2023ApJSong}.
Although the Type~3 FRB sample remains limited, it substantially improves the constraints on $\rm DM_{host}$, from $(\mu_{\rm host}=2.98^{+1.18}_{-1.94}, \, \sigma_{\rm host}=1.43^{+0.78}_{-0.70})$ to $(\mu_{\rm host}=4.08^{+0.56}_{-0.58}, \, \sigma_{\rm host}=0.90^{+0.52}_{-0.47})$. Furthermore, varying $\mathrm{DM}_{\rm halo}$ from $20$ to $50\,\mathrm{pc\,cm^{-3}}$ shifts the host-DM distribution from $(\mu_{\rm host},\sigma_{\rm host})\simeq(4.3,0.7)$ to $(2.3,1.6)$, but leaves $\Delta t_d$ within $1\sigma$ of its fiducial value. Crucially, the aforementioned concordance is not driven by a preference for any particular DTD model or Madau-Extended model. The log-Bayes factors \footnote{$\ln\mathcal{B}$ is the absolute difference in log Bayesian evidence $\ln\mathcal{Z}$ between two models \cite{1939Jeffreys}.} of all DTD models agree to within $\ln\mathcal{B}<1.5$, while their maximum log-likelihoods differ by less than unity, indicating that the current data cannot provide statistical preference among them. We notice that there is an obvious degeneracy between $\rm DM_{host}$ and $\rm DM_{halo}$, but our main conclusion is robust under varied assumptions of different DM models. In the Supplementary Materials, we provide the complete posterior distributions, detailed model comparisons, and an additional end-to-end validation to demonstrate the robustness.

All physically motivated DTD models converge on the same conclusion. The FRB event rate peaks at essentially the same redshift as the cosmic SFH when a specific SFH is assumed. The Madau-Extended model, which adopts a more flexible parametric form of $\mathcal{R}(z)$, also predicts a large redshift peak. We therefore conclude that at least the majority of the FRB population closely traces cosmic star formation with, at most, a short delay. Although the data do not favor any specific DTD parametrization, they robustly support a common physical picture in which FRBs are produced by progenitors with characteristic delay times well below the multi-Gyr timescales.

\section{Discussion} \label{sec:summary}
Our analysis shows that the cosmic FRB rate peaks at essentially the same redshift as the cosmic SFH, with a median characteristic delay $\Delta t_d\sim 0.1$--$0.3$ Gyr across all delay-time prescriptions, carrying a direct implication for the progenitor question. A natural benchmark is the DTD of compact binary mergers, which has been probed extensively through sGRBs but remains under active debate. Early phenomenological fits to the sGRB rate and luminosity function favored lognormal DTDs with typical delays of $3$--$4$~Gyr \cite{2007PhRNakar, 2015MNRASWanderman, 2016A&AGhirlanda}, whereas some recent hierarchical inferences favor a DTD steeper than the canonical $t^{-1}$ with shorter delay \cite{2022ApJZevin, 2022ApJNugent,2026A&APracchia} (see also \cite{2021arXiv210907694J} for the evidence for the prompt merger). Several joint fits still recover characteristic delays of $\sim 1$~Gyr \cite{2022MNRASLuo,2025MNRASGao}, and the sGRBs in old, quiescent hosts \cite{2022ApJFong}, such as the $\gtrsim 10$~Gyr-old stellar population hosting GW170817 \cite{2017ApJBlanchard}, suggest that $\gtrsim 20\%$ of mergers take place at delays $> 1$~Gyr \cite{2026arXiv26Chattaraj}. With delays of a few hundred Myr, binary white dwarf mergers could also be potential FRB progenitors \cite{2018ApJCaoXF}. These current constraints thus bracket the typical merger delay between a few hundred Myr and a few Gyr. The FRB population analyzed here excludes characteristic delays $\Delta t_d > 1$ Gyr at the 90\% credible level and requires a median delay of only $0.1$--$0.3$~Gyr, which is consistent with zero delay at the $2\sigma$ level and lies at or below the shortest end of this range.

Hence, at least for the bulk of the FRB sources, the formation channels likely differ from those of GW170817/GRB 170817A-like events with a merger timescale of $\sim 1-10$ Gyr \cite{2017ApJBlanchard}. The cosmic FRB rate follows the assumed SFH essentially instantaneously and exhibits a mild super-linear scaling with $n>1$. Such an observed super-linear trend is physically plausible if FRBs favor a low-metallicity channel. For example, at low metallicity, weaker line-driven winds allow massive stars to retain the rapid core rotation required for dynamo amplification of magnetar-strength fields \cite{1992ApJDuncan, 2005A&AYoon}, while the cosmic fraction of the low-metallicity SFR density rises from ${\sim}15\%$ at $z=0$ to ${\sim}40\%$ at $z=2$ \cite{2006ApJLangerNorman, 2019MNRASChruslinska}. This makes young magnetars or neutron stars formed in core-collapse supernovae in low-metallicity environments natural candidates \cite{2018ApJMargalit, 2021ApJBochenek}. The prompt neutron star mergers, in principle, could also contribute.

When compared with previous population analyses, we note that the reconstructions employing the Lynden-Bell $c^-$ method or empirical selection functions of fluence consistently report rates that track the sGRB rate with substantial delays \cite{2022JCAPQiangDC, 2022ApJZhangRachel, 2024ApJChenJH, 2024ApJLinHN, 2025ApJChampati, 2025A&AZhangKJ, 2025ApJZhouH, 2026arXiv26SuZW}. 
For the CHIME/FRB catalog, because the beam response varies significantly across the header localization region of each burst, only a lower limit on the fluence is provided. Methods that rely on a simplified flux-limited truncation can therefore suppress the inferred high-redshift population and bias the rate peak toward lower redshift. Similar biases have recently been reported in sGRB population analyses \cite{2026A&APracchia, 2026arXivSantis}, which argue that the longer time delays inferred in earlier studies most likely stem from an incorrect treatment of selection effects. By combining forward modeling with hierarchical Bayesian inference, our analysis removes this bias and recovers a rate peak coincident with cosmic star formation.

These conclusions characterize chiefly the one-off FRB population, since only nine repeaters are included. Any difference in the population parameters between analyses with and without repeating sources is not statistically distinguishable. Our inference of a dominant prompt channel also leaves FRBs in old stellar environments, such as the repeater FRB~20200120E, unexplained, possibly suggesting a subdominant delayed channel not captured by our results. Future analyses can be sharpened by several developments: (i) more localized FRBs with spectroscopic redshifts from MeerKAT, DSA-110, CHIME/FRB Outriggers and ASKAP, which would break the DM--$z$ degeneracy model-independently; (ii) higher-redshift localized bursts to sample the peak of $\mathcal{R}(z)$ directly; (iii) multi-frequency detections to constrain the energy--fluence--redshift conversion and test the assumed spectral index and rest-frame bandwidth, particularly at low frequencies given the chromaticity of FRB activity \cite{2021ApJPleunis, 2021NaturMarazuela}; (iv) improved $\mathrm{DM}_{\rm host}$ priors and fits for the feedback parameter $\mathcal{F}$; (v) separate population models for repeating and one-off FRBs. Taken together, these developments may identify the preferred DTD and reveal any difference in $\mathcal{R}(z)$ between repeaters and one-off FRBs. \\

\Acknowledgements{
This work is supported in part by NSFC under grants of No.~12588101, No.~12233011, and No.~12503059, the Strategic Priority Research Program of the Chinese Academy of Sciences (grant No.~XDB0550400), the Postdoctoral Fellowship Program of CPSF (No.~GZB20250738), the China Postdoctoral Science Foundation (grant No.~2025M783236), and the Jiangsu Funding Program for Excellent Postdoctoral Talent (No.~2025ZB209).}

\InterestConflict{The authors declare that they have no conflict of interest.}

\bibliography{reference}{}
\bibliographystyle{scpma.bst}



\end{multicols}
\end{document}


\title{\textit{Supplemental Material}: Fast Radio Bursts Trace Cosmic Star Formation\\ with Little Delay}

\maketitle

\setcounter{section}{0}
\renewcommand\thesection{S\arabic{section}}
\setcounter{figure}{0}
\renewcommand\thefigure{S\arabic{figure}}
\setcounter{table}{0}
\renewcommand\thetable{S\arabic{table}}

\section{Sample Selections}\label{sec:selection}
Following Ref.~\cite{2023ApJShin}, we select 225 sources from CHIME/FRB Catalog~1 \cite{2021ApJSCHIME}, retaining only the first detected burst that satisfies the selection criteria from each repeater (9 repeaters are included). The total observing time is $T_{\rm obs} = 214.8$ days, and the total number of sources expected during CHIME/FRB catalog 1 is $N=\int{\Phi(z|{\bf \Lambda})\frac{{\rm d}V}{{\rm d}z}\frac{T_{\rm obs}}{1+z}{\rm d}z}$, where $\Phi(z|\Lambda_3) = \Phi_0 \frac{\mathcal{R}(z|\Lambda_3)}{\mathcal{R}(0)}$ is the FRB volumetric event rate. A detected source is retained if it satisfies all of the following criteria:
\begin{itemize}
    \item ${\rm S/N}_{\rm bonsai} \ge 12$;
    \item a fitted dispersion measure ${\rm DM}_{\rm fitb} \ge 100\,\mathrm{pc\,cm^{-3}}$;
    \item a limited Galactic contribution, $\rm DM_{ISM}^{NE2001} < 100\, \mathrm{pc\,cm^{-3}}$;
    \item an extragalactic-DM dominance condition, ${\rm DM}_{\rm fitb} > 1.5\, \max[\rm DM_{ISM}^{NE2001}, \, DM_{ISM}^{YMW16}]$, where the Galactic contributions are predicted by the Milky Way electron density models NE2001 \cite{2002astroCordes} and YMW16 \cite{2017ApJYJM};
    \item a scattering timescale $\tau_{\rm scat} \le 10\,\mathrm{ms}$ at 600 MHz;
    \item a clean detection, requiring \texttt{excluded\_flag}$=0$ and 
    \texttt{sub\_num}$=0$ (i.e. retaining only the primary, non-flagged sub-burst).
\end{itemize}
In addition, bursts detected during pre-commissioning, during epochs of low sensitivity, or on days with software upgrades are excluded.

For the injection samples, $N_{\rm inj} = 5\times10^6$ is the total number of injections, $N_{\rm found}=14514$ is the number of injections that satisfy the selection criteria, and $P_{\rm inj}$ is the probability distribution from which the injections are drawn. Specifically, $P_{\rm inj}(\rm DM)$ is a mixture of log-normal (90\%) and uniform (10\%) components, while $P_{\rm inj}(F)$ follows a power-law distribution with index $\gamma = -1$. We apply the same selection criteria to the injections as to the observations, additionally requiring a beam position within the central region ($-5 \le x_{\rm beam} \le 5$) and either a radio-frequency-interference (RFI) grade above threshold ($G_{\rm RFI} > 7$) or a sufficiently high S/N override (${\rm S/N}_{\rm bonsai} > 30$), so that high-significance events are retained regardless of their RFI grade. Accounting for the injection efficiency $\epsilon_{\rm inj}=0.874$, the expected number of detections becomes $N_{\rm exp} = N\xi({\bf \Lambda})\epsilon_{\rm inj}$.

\section{Population Models}\label{sec:model}
Because nearly all FRBs detected by CHIME are not precisely localized, their redshift distributions need to be inferred jointly from the fluence $F$ (brightness) and DM (distance). Other properties, such as intrinsic width and scattering times, are not used because of their weak correlation with redshift. The joint distribution of $F$ and DM is
\begin{equation}\label{eq:1}
\pi(F, {\rm DM}|{\bf \Lambda}) = \int {\rm d}z \pi(F|z,\Lambda_1) \pi({\rm DM}|z,\Lambda_2)\pi(z|\Lambda_3),
\end{equation}
where $\pi(z|\Lambda_3)$ follows Eq.~(7) of the main text and each term is normalized. Assuming that the FRB energy distribution does not evolve with redshift (see \autoref{sec:E-Frelation} for more details), it can be described by the Schechter function with an exponential cutoff
\begin{equation}
P(E) \propto \frac{1}{E_{\rm char}} \bigg(\frac{E}{E_{\rm char}} \bigg)^{\gamma} \exp{\bigg(-\frac{E}{E_{\rm char}} \bigg)}, \, \, E>E_{\rm pivot},
\end{equation}
where $E_{\rm char}$ is the characteristic cutoff energy, and $E_{\rm pivot}$ is the minimum energy of the modeled population. Assuming $\alpha$ is the FRB spectral index, the fluence behaves as $F\sim \nu^{\alpha}$ and the resulting fluence distribution is
\begin{equation}
\pi(F|z, \Lambda_1) \propto P(E(F, z))\bigg|\frac{{\rm d}E}{{\rm d}F} \bigg|, 
\end{equation}
with $\frac{{\rm d}E}{{\rm d}F} = 4\pi d_{L}^2(z)\Delta \nu (1+z)^{-(2+\alpha)}$, where $\Delta \nu = 1 \rm \, GHz$ is the frequency bandwidth and $\Lambda_1 = \{E_{\rm char}, \, \gamma, \, \alpha \}$. 

The observed dispersion measure is decomposed as $\mathrm{DM_{obs}} = \mathrm{DM_{halo}} + \mathrm{DM_{ISM}} + \mathrm{DM_{host}} + \mathrm{DM_{cosmic}}$, comprising contributions from the Milky Way halo and disk, the host galaxies, and the diffuse cosmic baryons. Because the estimate of $\rm DM_{halo}$ remains highly uncertain, ranging from $10-80 \rm \, pc \, cm^{-3}$ \cite{2015MNRASDolag, 2019MNRASProchaska, 2020ApJYamasaki, 2020MNRASKeating,2026arXiv26ZhangYC}, we adopt a fixed $\rm DM_{halo} = 30\rm \, pc \, cm^{-3}$ and calculate $\rm DM_{ISM}$ with the NE2001 model \cite{2002astroCordes} as the fiducial values. Previous studies have shown that the assumed $\rm DM_{halo}$ has little impact on the inferred $\rm DM_{cosmic}$, since the contributions can be absorbed between the halo and host components while yielding a consistent cosmic-DM estimate \cite{2025NatAsConnor}. The distribution of $\rm DM_{cosmic}$ is approximated by a quasi-Gaussian function with an extended tail \cite{2020NaturMacquart}, 
\begin{equation}
P_{\rm cosmic}(\Delta)=A\Delta^{-\beta_{\rm M}}{\rm exp} \left[ -\frac{(\Delta^{-\alpha_{\rm M}} - C_0)^2}{2\alpha_{\rm M}^2\sigma_{\rm DM}^2}\right], \quad \Delta>0,
\end{equation}
where $\Delta = {{\rm DM}_{\rm cosmic}}/\langle {\rm DM}_{\rm cosmic} \rangle$, $A$ is the normalization factor, $\alpha_{\rm M}=\beta_{\rm M}=3$ are the inner and outer slopes of the gas density profile. The effective standard deviation is $\sigma_{\rm DM} = \mathcal{F}z^{-0.5}$, with the feedback parameter $\mathcal{F} = 0.32$ \cite{2022MNRASJames}, and $C_0$ is fixed by requiring $\langle \Delta \rangle =1$. The mean cosmic $\rm DM_{cosmic}$ is given by \cite{2012ApJShull, 2014ApJDeng}
\begin{equation}\label{eq:DM_IGM}
\langle {\rm DM}_{\rm cosmic} \rangle (z)= \frac{21 c \Omega_b H^2_0 f_{\rm d}}{64\pi G m_p}\int^z_0 \frac{1+z'}{H(z')}{\rm d}z'.
\end{equation}
Because the cosmic DM includes contributions from both the intergalactic medium and intervening halos, which are described by similar formalisms, we adopt a single diffuse ionized baryon fraction $f_{\rm d} = 0.94$ for simplicity \cite{2025NatAsConnor}. The remaining cosmological parameters are fixed to the Planck 2018 results \cite{2020A&APlanck}. The observed $\rm DM_{cosmic}$ distribution then reads \cite{2023ApJShin, 2025PhRvDZhang, 2026ApJZhuge}
\begin{equation}
P({\rm DM_{cosmic}|} z) = P_{\rm cosmic}(\Delta) \frac{1}{\langle {\rm DM}_{\rm cosmic} \rangle (z)}.
\end{equation}
Thus, only $\rm DM_{host}$ is treated as free and is modeled with a log-normal distribution,
\begin{equation}\label{eq:host}
\begin{aligned}
P({\rm DM_{host}|\Lambda_2}) &=  \frac{1}{\sqrt{2\pi}{\rm DM_{host}}\sigma_{\rm host}} \\
& \times {\rm exp} \left[ - \frac{({\rm ln \, DM}_{\rm host} - \mu_{\rm host})^2}{2\sigma_{\rm host}^2} \right],
\end{aligned}
\end{equation}
where $\Lambda_2 = \{ \sigma_{\rm host}, \, \mu_{\rm host}\}$. In the observer frame, the host contribution is related through ${\rm DM_{host}^{obs}} = {\rm DM_{host}^{src}}/(1+z)$, so that
\begin{equation}
P({\rm DM_{host}^{obs}}|\Lambda_2,z) = P({\rm DM_{host}^{src}}|\Lambda_2)(1+z).
\end{equation}
Since the Milky Way term $\rm DM_{MW} = DM_{ISM}+DM_{halo}$ is constant, the distribution of $\rm DM_{obs}$ becomes
\begin{equation}
\begin{aligned}
&\pi({{\rm DM_{obs}|\Lambda_2},z}) \\
&\propto \int P({\rm DM_{host}^{src}}|\Lambda_2)P({\rm DM_{cosmic}|} z)(1+z)\rm d DM_{cosmic},
\end{aligned}
\end{equation}
with ${\rm DM_{host}^{src}} = (1+z)(\rm DM_{obs} - DM_{MW} - DM_{cosmic})$.

\begin{table*}[!tp]
\caption{\label{tab:table1}
The Prior Distributions of the Hyper Parameters}
\begin{tabular}{lcc}
\hline
Descriptions &Parameters &Priors\footnotemark[1]\\
\hline
Local volumetric rate above $E_{\rm pivot}$ &$\log_{10}\Phi_0$($\rm Gpc^{-3} \, yr^{-1}$) &U(-0.96, 6.43) \\
Differential power-law index of energy distribution &$\gamma$ &U(-2.5, 2.0) \\
Characteristic exponential cutoff energy &$E_{\rm char}$(erg) &U(38.00, 49.00) \\
Spectral index of $F$ &$\alpha$ &U(-5, 5) \\
Central value of the host DM distribution &$\mu_{\rm host}$ &U(0.001, 7) \\
Width of the host DM distribution &$\sigma_{\rm host}$ &U(0.001, 3) \\
Power-law index of the SFR scaling &$n$ &U(-2, 8) \\ 
\hline
Time delay of the fixed delay model &$\log_{10}t_d$(Gyr) &U(-3, 1) \\
\hline
Minimum time delay of the power-law delay model &$\log_{10}t_{d}^{\rm min}$(Gyr) &U(-3, 1) \\
Maximum time delay of the power-law delay model &$t_d^{\rm max}$(Gyr) &U(3, 13.8) \\
Index of the power-law delay model &$\beta$ &U(-3, 1) \\
\hline
Central value of the log-normal delay model &$\mu_{t_d}$ &U(-6.9, 2.3) \\
Width of the log-normal delay model &$\sigma_{t_d}$ &U(0.01, 10.0) \\
\hline 
Index of the Madau-extended model &$\lambda$\footnotemark[2] &U(0.001, 6)\\
Index of the Madau-extended model &$\kappa$\footnotemark[2] &U(2, 10) \\
Redshift peak of the Madau-extended model  &$z_p$ &U(0.001, 4) \\
\hline

\end{tabular}
\footnotetext[1]{Here, ``U" represents the uniform distribution}
\footnotetext[2]{$\kappa > \lambda$ is required to ensure the finite integral of the event rate.}
\end{table*}

\begin{table*}[!tp]
\caption{\label{tab:table2}
Model Comparison Results}
\begin{tabular}{lccccc}
\hline
Models &$\ln \rm (likelihood_{\rm max})$ &$\ln \mathcal{Z}$ &$\ln \mathcal{B}$\footnotemark[1] &$N_{\rm eff}$ &$N_{\rm eff, min}$\\
\hline
Power-law Delay &-1285.04 & -1301.01 &0.72 &$11521.48^{+493.02}_{-514.79}$ &9296.20\\
Log-normal Delay &-1284.76 &-1300.32 &1.41 &$10771.65^{+695.68}_{-754.81}$ &7374.27\\
Fixed Delay &-1285.07 &-1301.54 &0.19 &$11026.48^{+637.04}_{-663.17}$ &8348.08\\
SFH tracking (${\rm SFR}^n$) &-1285.17  &-1301.73 &0 &$10997.87^{+629.84}_{-664.51}$ &8431.29\\ 
Madau extended &-1284.82 &-1300.60 &1.13 &$11016.06^{+595.04}_{-644.70}$ &8376.30\\
\hline
Power-law Delay ($n=1$) &-1285.60 &-1301.13 &0.60 &$11620.86^{+485.58}_{-510.17}$ &9656.85\\ 
Power-law Delay (3 pars) &-1285.08 &-1301.28 &0.45 &$11090.19_{-681.04}^{+601.64}$ &7785.75\\ 
\hline
\end{tabular}
\footnotetext[1]{$\ln \mathcal{B}$ is referenced to the SFH-tracking model. Evidence is commonly graded as $\ln\mathcal{B}=0$--$1.2$ (inconclusive), $1.2$--$2.3$ (substantial), $2.3$--$4.6$ (strong), and $>4.6$ (decisive).}
\end{table*}

The priors of all the parameters are summarized in Table~\ref{tab:table1}. To balance the efficiency and accuracy, we employ the nested sampling algorithm implemented in {\tt pymultinest} \cite{2016pymultinest} and set 1000 live points during Bayesian analysis. 
We therefore evaluate $N_{\rm eff}$ for every hyperparameter posterior sample of each population model shown in Figure~2 of the main text and \autoref{fig:PL_rate}. The results are summarized in Table~\ref{tab:table2}. The global minimum over all posterior samples and models is $N_{\rm eff,min}>7000$, compared with $4N_{\rm det}=900$ for $N_{\rm det}=225$ observed FRBs in our analysis. This indicates that the injection sample provides adequate importance-sampling coverage of the detectable population across the posterior, and that the selection-integral estimate is not dominated by a small number of injections with disproportionately large weights.

\section{Selection effects}
\begin{figure}
\centering
\includegraphics[width=0.95\linewidth]{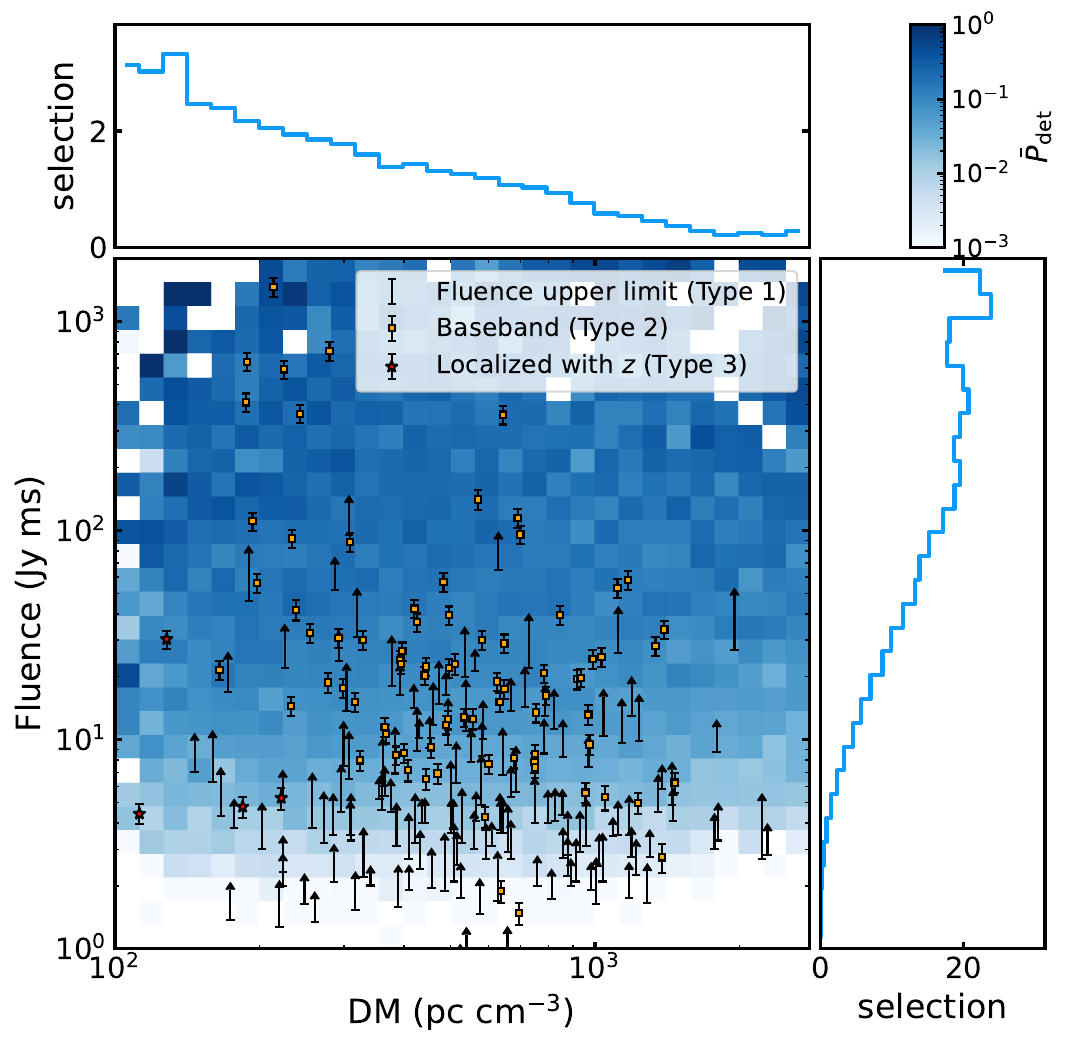}
\caption{Empirical CHIME/FRB selection under the best-fit population model. The main panel shows the binned detection-and-sample-inclusion probability $\overline{P}_{\rm det}({\rm DM},F)$. Black arrows, orange squares, and red stars mark the observed events. The side panels present the relative marginal selection functions $s({\rm DM})/\langle s\rangle$ and $s(F)/\langle s\rangle$.}
\label{fig:MySF}
\end{figure}

\begin{figure}
\centering
\includegraphics[width=0.95\linewidth]{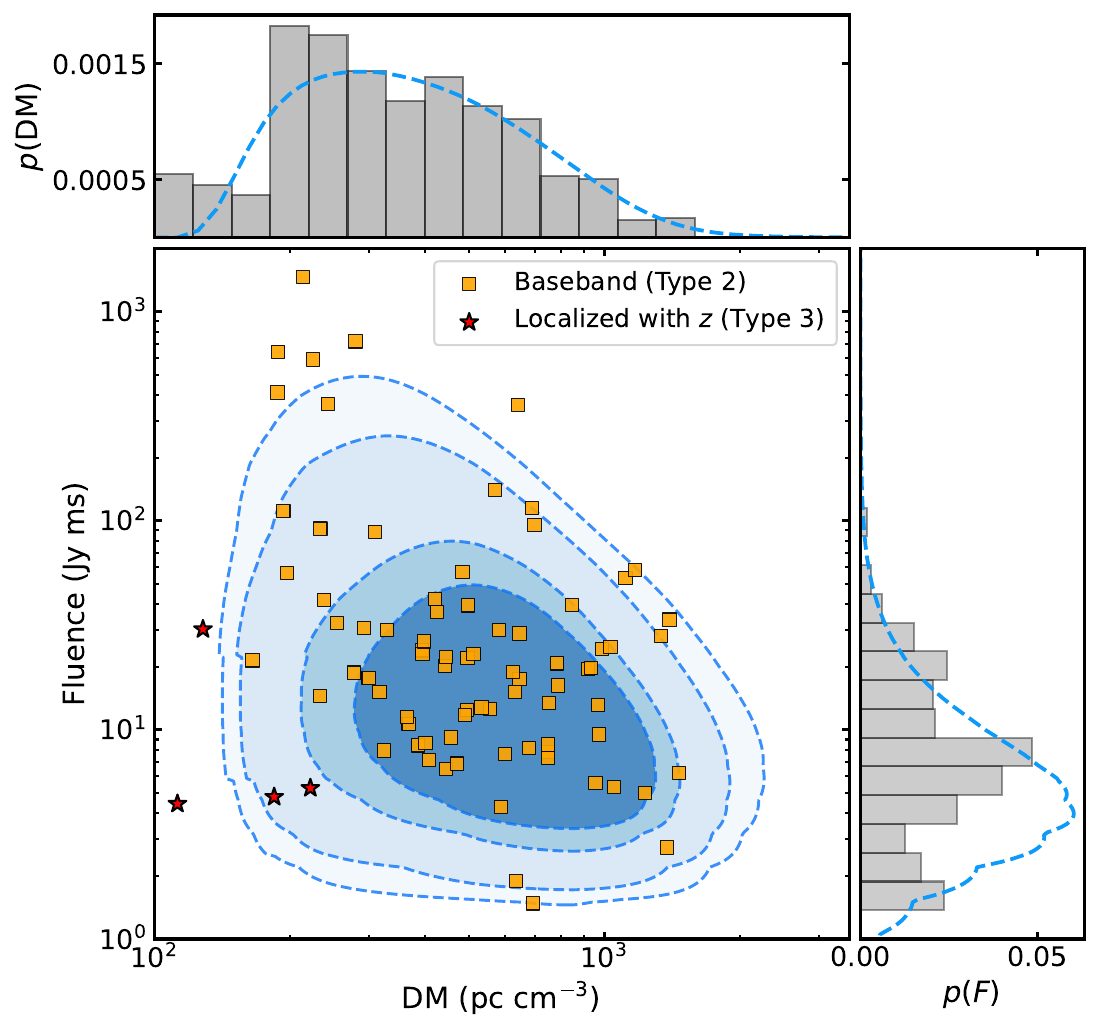}
\caption{Model-predicted density of detected events in the fluence–DM plane, compared with the baseband sample (Type 2, orange squares) and the localized events with measured redshifts (Type 3, red stars). Blue contours enclose 50\%, 68\%, 90\%, and 95\% of the detection density. }
\label{fig:DMF}
\end{figure}

In our analysis, the selection efficiency enters the likelihood function through the detection efficiency $\xi({\bf \Lambda})$, which is recomputed for every hyperparameter sample from the full injection set. No factorization of the instrumental response into independent fluence and DM terms is made. As defined in \cite{2021ApJSCHIME}, the intrinsic distribution of DM can be corrected by the observed distribution
\begin{equation}
P({\rm DM}) = P_{\rm obs}({\rm DM})s(\rm DM)^{-1},
\end{equation}
where $s({\rm DM})$ is the DM selection function marginalized over other FRB properties. Corresponding to Figure~1 of the main text, we calculate the weights $w_j$ (Eq.~(4) of the main text) of detected injections and all injections at the best-fit value. The detection probability in a $(\mathrm{DM},F)$ bin is estimated as ${\overline{P}}_{\rm det}(\mathrm{DM},F)=\frac{\sum_{j\in{\rm detected}} w_j}{\sum_{j\in{\rm all\,inj}} w_j}$. This quantity is the detection probability marginalized over the remaining burst properties under the injection population. As shown in \autoref{fig:MySF}, the corresponding one-dimensional curves are obtained by applying the same weighted ratio in DM or fluence bins. The one-dimensional selection functions show that bright bursts are more likely to be detected. High-DM bursts are distant and faint, and only the luminous tail survives the fluence threshold. The impact of this selection on the observable plane is illustrated in \autoref{fig:DMF}, which shows the detected fluence–DM density predicted by the best-fit population model with no delay assumption. Here, the intrinsic population propagates through the selection function and convolves with the fluence measurement uncertainty. The DM and fluence marginal distributions give $p_{\rm PPC}=0.57$ and $p_{\rm PPC}=0.11$, respectively. The combined 2D distribution gives $p_{\rm PPC} = 0.11$. The agreement between the contours and the observed baseband sample provides a visual validation of both the population model and the observations.

\begin{figure}
\centering
\includegraphics[width=0.95\linewidth]{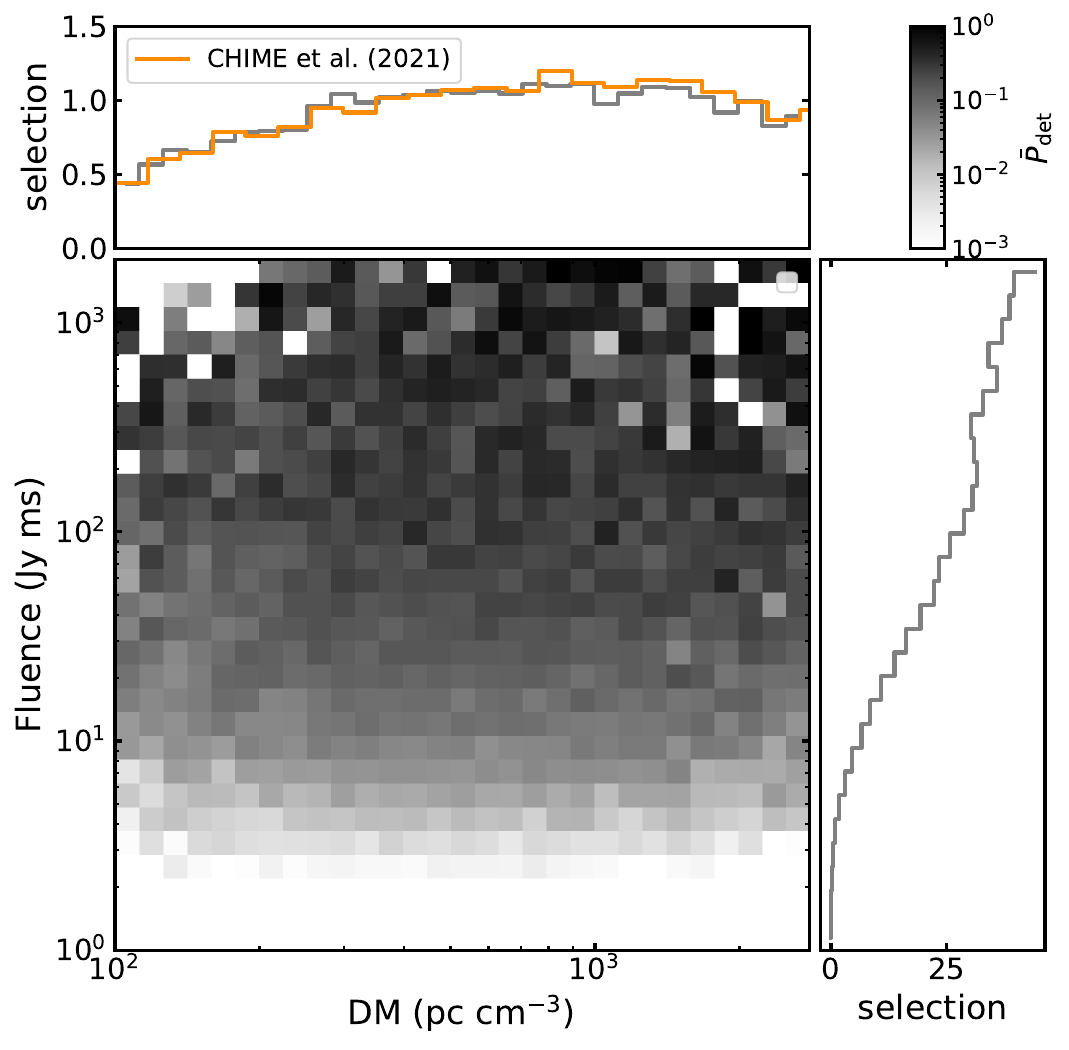}
\caption{Selection function compared with \cite{2021ApJSCHIME}. The central panel shows the estimated detection probability $P(\rm DM, F)$. The top panel shows the marginalized relative selection function as a function of DM. The blue curve is our reconstruction, while the orange curve is the result reported by \cite{2021ApJSCHIME}. Both of the one-dimensional curves use the same fiducial-population normalization, such that they represent relative selection functions rather than probabilities bounded by unity. The right panel shows the corresponding marginalized fluence selection function from our weighted injection sample.}
\label{fig:SF} 
\end{figure}

For a direct comparison with the selection functions of Ref. \cite{2021ApJSCHIME}, the same estimator is evaluated under their fiducial population. We define $w_j^{\rm fid}=\frac{p_{\rm fid}(\theta_j)}{P_{\rm inj}(\theta_j)}$ and use the fiducial distributions adopted for the CHIME comparison, including uncorrelated log-normal distributions for DM ($\sigma=0.66$, scale $495\,{\rm pc\,cm^{-3}}$), pulse width ($\sigma=0.97$, scale $1\,{\rm ms}$), and scattering time ($\sigma=1.72$, scale $2.02\,{\rm ms}$), together with a power-law fluence distribution $p_{\rm fid}(F)\propto F^{\alpha-1}$ with $\alpha=-1.32$ and pivot $F_0=0.2\,{\rm Jy\,ms}$.

Following the normalization convention of \cite{2021ApJSCHIME}, each one-dimensional curve is divided by its fiducial-population-averaged efficiency. Thus, the curves shown in \autoref{fig:SF} are relative marginal selection functions and are not probabilities bounded by unity. They are used only as a validation of the injection calculation, and the inference itself uses the full injection-weighted efficiency $\hat{\xi}({\bf \Lambda})$ for each hyperparameter sample. The DM-dependent curve reconstructed from the injections closely follows the CHIME/FRB Catalog 1 curve over the common DM range. We emphasize that the binned maps and curves shown here are diagnostics. None of them enters the inference. The likelihood uses the full-resolution per-event importance sum of the detection efficiency.

\section{Detailed Results and Model comparison}\label{sec:compare}

\begin{figure}
\centering
\includegraphics[width=0.95\linewidth]{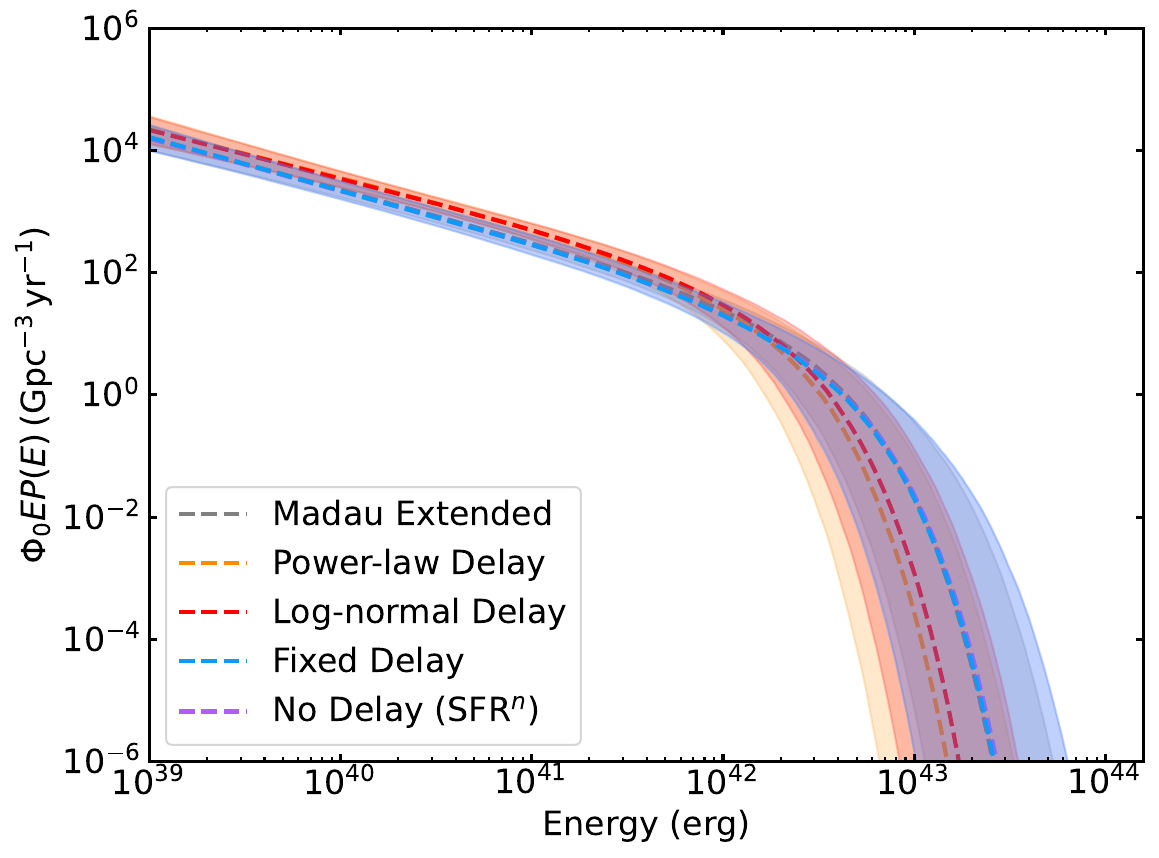}
\caption{The posterior Schechter function results scaled by energy and the volumetric rate, corresponding to the same scenarios presented in Figure~2 of the main text. The shaded regions represent the 68\% credible intervals, while the dashed lines indicate the median values.}
\label{fig:energy} 
\end{figure}

The full posterior distributions of the models shown in Figure~2 of the main text are presented in \autoref{fig:posa} and \autoref{fig:posb}. The two figures share a common set of seven parameters that describe the FRB energy function, the host-galaxy DM contribution, and the local volumetric rate. These population parameters are remarkably stable across all models/scenarios. They yield $\log_{10}\Phi_0 \simeq 5.1$–$5.2$, $\gamma \simeq -1.8$, $\log_{10}E_{\rm char}\simeq 42$, $\alpha \simeq -3$ with $\mu_{\rm host}\simeq 4.5$ and $\sigma_{\rm host}\simeq 1.0$, in broad agreement with previous studies \cite{2020MNRASLuo, 2023ApJShin, 2025NatAsConnor}. We note that the population-level log-normal $\mathrm{DM_{host}}$ is a simplification, and that individual sources can depart from it dramatically. The most extreme example is FRB~20190520B, with $\mathrm{DM_{host}}\simeq903$~pc~cm$^{-3}$, nearly an order of magnitude above the host average \cite{2022NaturNiu, 2022ApJOcker}. Although this source is not in our Catalog~1 sample, under our median posterior, the probability of such a large host contribution is $P(\rm DM_{host} > 900 \, pc \,cm^{-3}) = 1.2\times 10^{-3}$, corresponding to $\sim 0.3$ sources expected in a sample of $225$ events. Our model therefore accommodates such extreme sources but predicts them to be rare. As shown in \autoref{fig:energy}, the energy distributions derived from different DTD models are well constrained and exhibit highly consistent shapes with one another. The insensitivity of these parameters to the assumed parametrization of $\mathcal{R}(z)$ demonstrates that our constraints on the FRB energy distribution and host environment are internally self-consistenct.

\begin{table}[!htbp]
\caption{\label{tab:bw}
Bandwidth dependence of the Type 1 KDE}
\begin{tabular}{lcc}
\hline
h &$\Delta t_d \, (\rm Gyr)$ &$z_{\rm peak}$ \\
\hline
0.05 &$0.09^{+0.41}_{-0.09}$ &$1.83^{+0.05}_{-0.20}$\\
0.1 &$0.09^{+0.41}_{-0.09}$ &$1.83^{+0.05}_{-0.20}$\\
0.2 &$0.09^{+0.41}_{-0.09}$ &$1.83^{+0.05}_{-0.20}$\\
0.3 &$0.09^{+0.64}_{-0.09}$ &$1.83^{+0.05}_{-0.25}$\\ 
0.4 &$0.09^{+0.64}_{-0.09}$ &$1.83^{+0.05}_{-0.25}$\\
\hline
\end{tabular}
\end{table}

\subsection{The effect of the KDE bandwidth}
To check the robustness of our results to the choice of smoothing parameter in the 2D KDE over $\log \rm S/N$ and $\log \rm DM$ for Type~1 sources, we vary the bandwidth over $h\in[0.05, 0.4]$ and analyze the data with the fixed delay model. The resulting posterior constraints on $\Delta t_d$ and $z_{\rm peak}$ are summarized in Table~\ref{tab:bw}. We find that the median values of all model parameters are unchanged to within $<1\sigma$, while the posterior widths increase monotonically with $h$ as expected. This confirms that our conclusions are robust to the choice of $h$, since $h$ describes the smoothness and only affects the effective resolution of the injection density.

\subsection{$E-F$ relation and energy function}\label{sec:E-Frelation}
\begin{figure}
\centering
\includegraphics[width=0.95\linewidth]{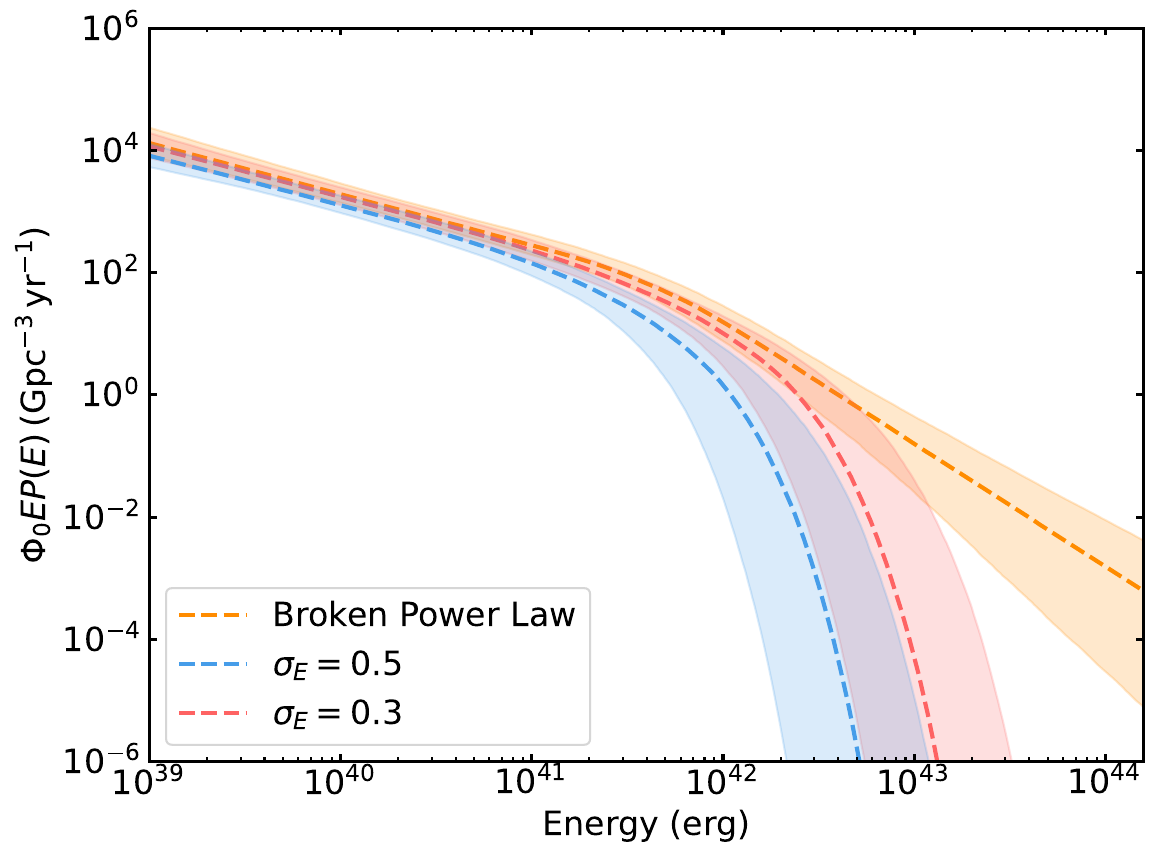}
\caption{The posterior Schechter and Broken Power-Law functions results with intrinsic scatter in the $E$--$F$ relation, $\sigma_E=0.3$ and $0.5$~dex.}
\label{fig:energy_sigE} 
\end{figure}

The preceding analysis maps fluence to energy deterministically with a single $\alpha$, and therefore does not capture the fine-scale spectral modulation \cite{2019ApJMacquart} or the band-limited morphology \cite{2021ApJPleunis} reported in the observed population. To account for the diversity of the FRB population, we introduce a multiplicative log-normal scatter of width $\sigma_E$ in the $E$--$F$ relation, $E_{\rm new}=E\,10^{s}$ with $s\sim\mathcal{N}(0,\sigma_E^{2})$, implemented as a convolution of the energy distribution in $\log_{10}E$. Because $\sigma_E$ and $E_{\rm char}$ both smooth the exponential cutoff and are therefore anti-correlated, we analyze fixed values $\sigma_E=0.3$ and $0.5$~dex. Between these two cases, $\log_{10}\Phi_0$, $\log_{10}E_{\rm char}$, and $\gamma$ shift from $4.98^{+0.27}_{-0.25}$, $41.93\pm0.43$, and $-1.82^{+0.15}_{-0.14}$ to $4.81^{+0.25}_{-0.24}$, $41.50^{+0.41}_{-0.42}$, and $-1.79^{+0.19}_{-0.16}$, respectively. The faint-end slope $\gamma$ is essentially unchanged, as expected. Log-normal scatter leaves the power-law slope invariant and rescales the normalization by $\exp[(\ln10)^{2}\sigma_E^{2}(\gamma+1)^{2}/2]$, while it only smooths the exponential cutoff and biases $E_{\rm char}$. The posteriors are shown in \autoref{fig:energy_sigE}.

For the intrinsic energy function we adopt a Schechter form with an exponential cutoff, which is the minimal function that combines a power-law rise with a high-energy turn-off. Such an exponential cutoff is preferred for three reasons. (i) It introduces only one characteristic scale, $E_{\rm char}$, in addition to the power-law slope $\gamma$. (ii) The exponential decay yields a convergent total rate and a finite mean energy without an artificial truncation. (iii) It represents a smooth suppression above a characteristic energy, a natural expectation if $E_{\rm char}$ is set by an energy reservoir or an efficiency limit of the emission mechanism, and it is the form widely used in earlier FRB population studies \cite{2018ApJFialkov, 2018MNRASLuo, 2021ApJNiu, 2023ApJShin}. To test the assumed redshift independence of the energy function, we allow $E_{\rm char}(z)=E_{\rm char,0}(1+z)^{\delta_E}$ with a broad prior $\delta_E\sim\mathcal{U}(-3,3)$ in the SFH-tracking model. We obtain $\gamma=-1.87^{+0.15}_{-0.13}$, $\log_{10}E_{\rm char}=42.23^{+0.53}_{-0.44}$, and $\delta_E=-0.24^{+1.86}_{-1.71}$. The posterior on $\delta_E$ is consistent with zero within its uncertainty, so the data show no evidence for evolution of the energy function with redshift. Such non-evolution results have also been reported in \cite{2023A&AZhangZL}. We emphasize that the single Schechter form is an approximation appropriate for the one-off-dominated sample considered here. It does not capture the bimodal burst-energy distribution reported for the well-sampled repeater FRB~121102\cite{2021NaturLi}, where FAST detected 1652 bursts following a combination of a log-normal function and a generalized Cauchy function.

Furthermore, we note that with the current data, the Schechter and broken power-law energy functions are not obviously distinguished. As shown in \autoref{fig:energy_sigE}, we assume the broken power-law model follows 
\[
P(E) \propto 
\begin{cases} 
\left(E/E_b\right)^{\gamma_{\rm lo}} & E_{\text{pivot}} \leq E < E_b \\ 
\left(E/E_b\right)^{\gamma_{\rm hi}} & E \geq E_b 
\end{cases}
\]
where $E_b$ is the break energy, $\gamma_{\rm lo}$ and $\gamma_{\rm hi}$ are the low- and high-energy slopes, and $\gamma_{\rm hi}<-1$ is required for convergence. With the SFH-tracking model and the priors $\log_{10}E_b\sim\mathcal{U}(38,\,43)$, $\gamma_{\rm lo}\sim\mathcal{U}(-2.5,\,0)$, and $\gamma_{\rm hi}\sim\mathcal{U}(-5,\,-1)$, we find $\log_{10}E_b=41.66^{+0.37}_{-0.34}$, $\gamma_{\rm lo}=-1.84\pm0.13$, and $\gamma_{\rm hi}=-3.00^{+0.35}_{-0.89}$. The Bayes factor between the broken power law and the Schechter model is $\mathcal{B}=1.42$. The two forms differ mainly in the high-energy tail, which the current data do not strongly constrain. We repeat the analysis with the fixed DTD model and find $\Delta t_d = 0.10^{+0.41}_{-0.09}$, suggesting that the delay time is not sensitive to the energy function formalism. Given this near-degeneracy, together with its parameter economy and its use in previous population studies, we retain the Schechter function as the fiducial energy function.

\subsection{Comparison between the prior and posterior}
\begin{figure}
\centering
\includegraphics[width=0.95\linewidth]{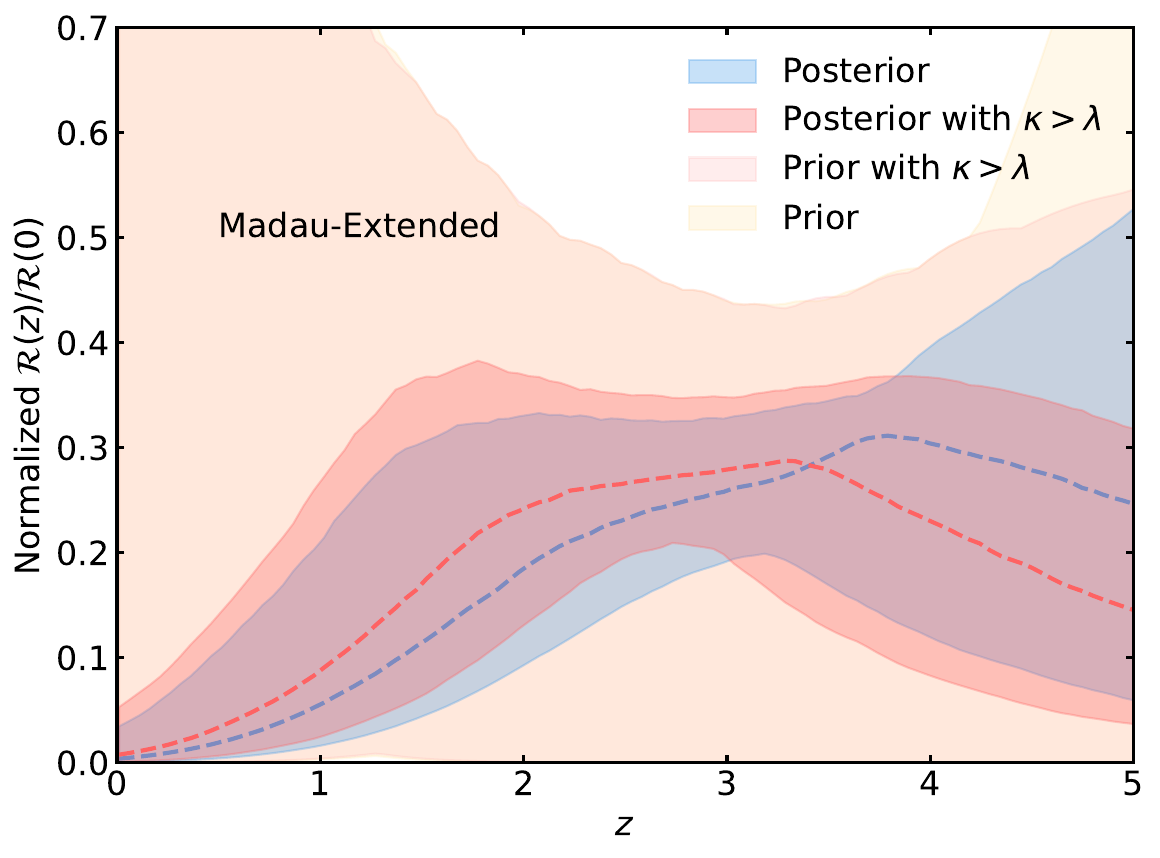}
\includegraphics[width=0.95\linewidth]{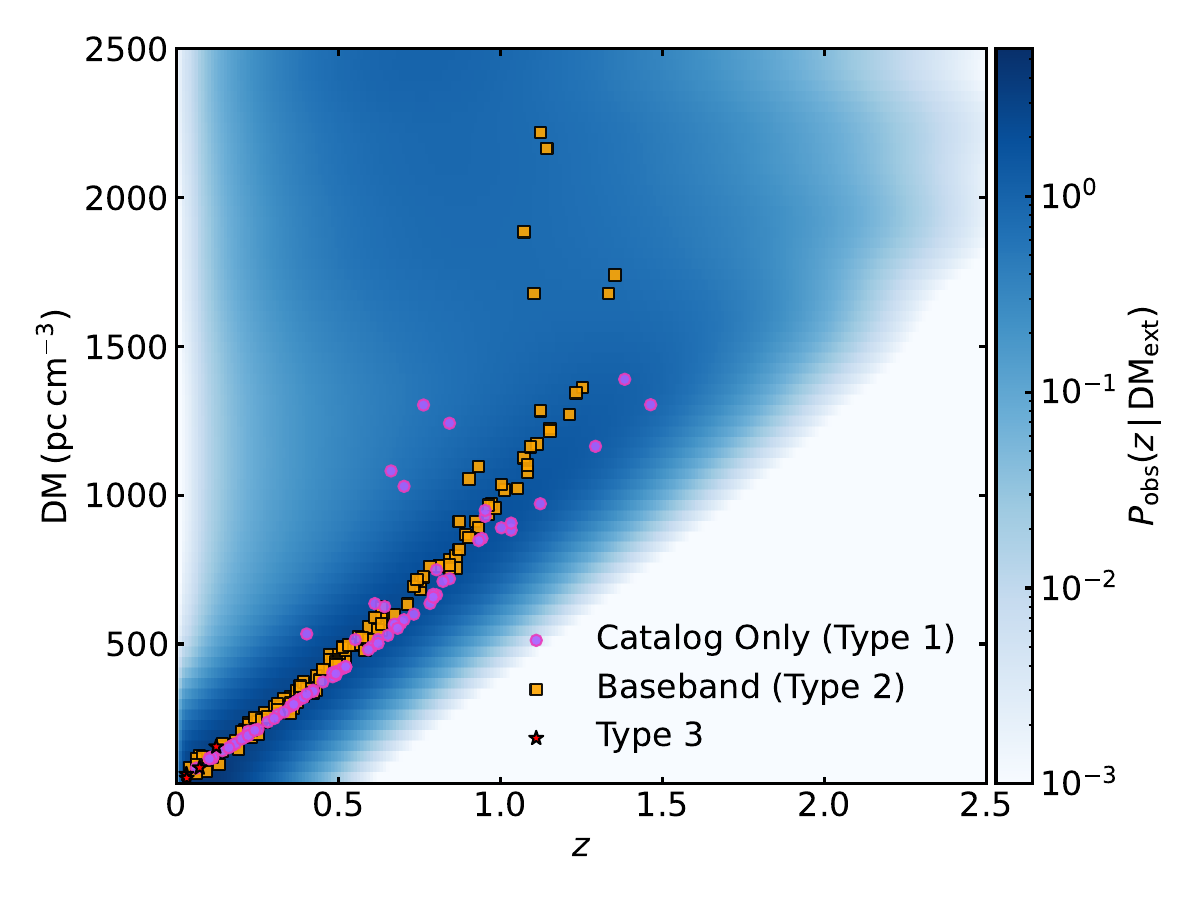}
\caption{Top: prior and posterior distributions of the normalized FRB rate $\mathcal{R}(z)/\mathcal{R}(0)$ for the Madau-Extended model. The shaded regions represent the 68\% credible intervals, and the dashed lines the medians. Bottom: redshift distributions of the sample under the best-fit values of the $\kappa>\lambda$ posterior. The background shows the normalized conditional distribution $P_{\rm obs}(z\,|\,\mathrm{DM_{ext}})$ (analogous to Figure~10 of \cite{2023ApJShin}), including the host and cosmic DM contributions as well as the fluence selection function. }
\label{fig:prior_pos} 
\end{figure}

As summarized in Table~\ref{tab:table2}, the maximum likelihood and Bayesian evidence of the different DTD models are statistically indistinguishable ($\Delta\ln\mathcal{Z}\lesssim 1$), indicating that the current CHIME/FRB Catalog~1 sample lacks sufficient statistical power to distinguish among competing DTD models. This limitation arises because the bulk of the detected bursts originate at low redshift ($z\lesssim 1$), where the predicted rate histories of the various models converge, leaving the high-redshift tail poorly sampled. To make explicit which features of $\mathcal{R}(z)$ are constrained by the data and which follow from the assumed functional form, we present the prior and posterior distributions of $\mathcal{R}(z)$, together with the redshift support of the sample in \autoref{fig:prior_pos}. The priors on the normalized rate are broad, and the posterior mainly constrains the low-redshift rise of $\mathcal{R}(z)$, which shows the feature that excludes long delay times. The bottom panel of \autoref{fig:prior_pos} quantifies the redshift range actually probed under the best-fit values of the Madau-Extended model, and the detections are marked at the maximum $P_{\rm obs}(z|\rm DM_{ext})$ values. Beyond the inferred peak of DTD models at $z_{\rm peak}\simeq1.8$, only $\sim0.4$ detections are expected. Consequently, the peak location of the DTD models and the declining branch of $\mathcal{R}(z)$ are not measured directly. At $z>3$ the apparent decline of the Madau-Extended posterior is driven by the convergence requirement $\kappa>\lambda$ imposed in the prior.

In the bottom panel of \autoref{fig:prior_pos}, the conditional distribution $P_{\rm obs}(z\,|\,\mathrm{DM_{ext}})$ exhibits a ``turnover'' at high DM, where the inferred redshift is much lower than predicted by a naive and roughly linear DM--$z$ relation. This behavior is fully consistent with \cite{2022MNRASJames, 2023ApJShin}. This behavior arises because a moderate redshift with a large $\mathrm{DM_{host}}$ contribution may be more probable than a high redshift with a large $\mathrm{DM_{IGM}}$ contribution, and this effect is amplified by the fluence selection against distant sources.

\begin{figure}
\centering
\includegraphics[width=0.95\linewidth]{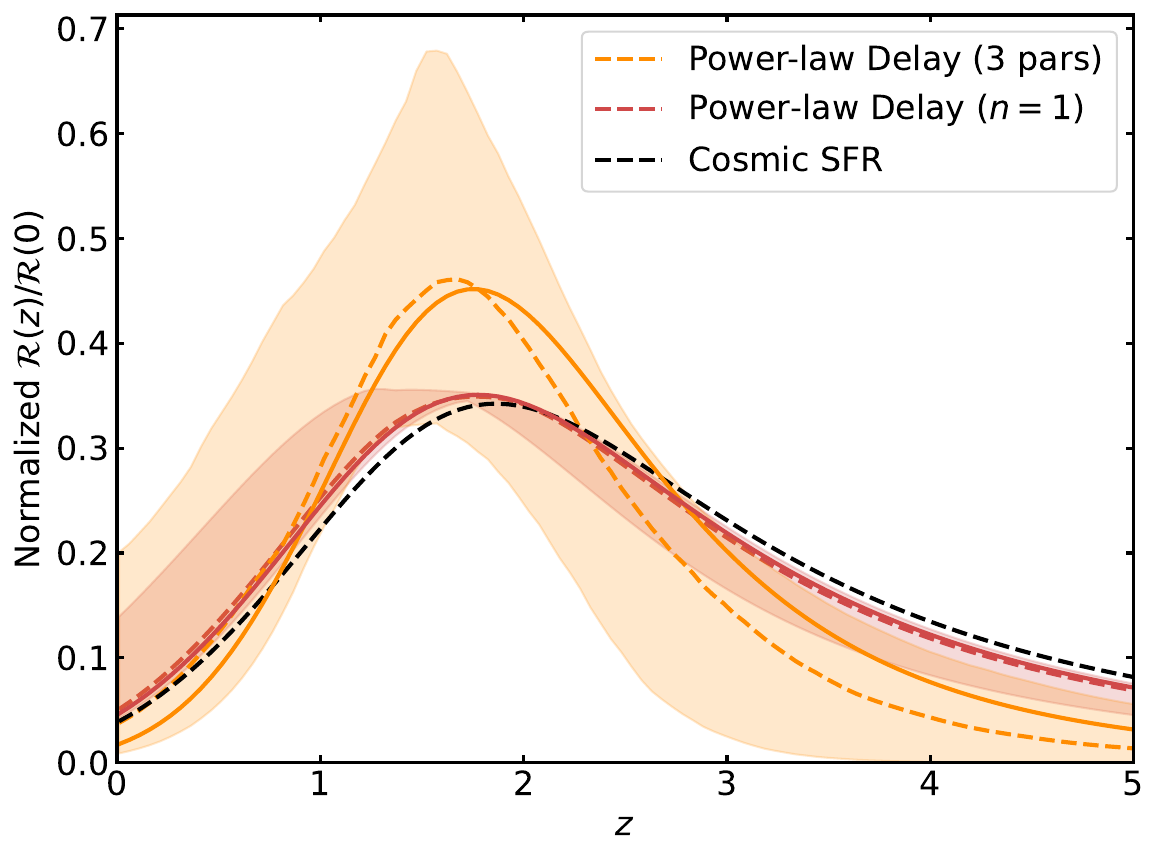}
\caption{The normalized intrinsic FRB event rate reconstructed for the power-law DTD models under different parametrizations. The three free parameters include $t_{\rm d}^{\rm min}$, $t_{\rm d}^{\rm max}$, and $\beta$. The shaded regions represent the 68\% credible intervals, while the dashed and solid lines indicate the medians and the best-fit values, respectively.}
\label{fig:PL_rate} 
\end{figure}

\subsection{Additional parameters of the Power-Law DTD}
To further assess the impact of the DTD parametrization, we examine two additional power-law delay scenarios. The reconstructed cosmic event rates are shown in \autoref{fig:PL_rate}. In the first case (``Power-law Delay, $n=1$'') we fix the rate density index to $n=1$, so that the FRB formation rate traces the cosmic star formation rate convolved with the DTD, adopting a minimum delay $t_{\rm d}^{\rm min}=10\,$Myr and a maximum delay $t_{\rm d}^{\rm max}=t_{\rm H}$.  In the second case (``Power-law Delay, 3 pars''), all three parameters $t_{\rm d}^{\rm min}$, $t_{\rm d}^{\rm max}$ and $\beta$ of the power-law model are allowed to vary. Both scenarios favor DTDs dominated by short delays, consistent with the other DTD models discussed above and supporting a population whose formation closely follows recent star formation, as expected if a substantial fraction of FRB progenitors are young magnetars born in core-collapse supernovae. The corresponding posterior distributions are compared in \autoref{fig:posc}.

\subsection{The effect of the localized FRBs}
The analyses above incorporate four localized FRBs, including FRB20181223C ($z=0.03024$), FRB20190110C ($z=0.12244$), FRB20190418A ($z=0.07132$), and FRB20190425A ($z=0.03122$) \cite{2024ApJBhardwaj, 2024ApJIbik}. These measured redshifts directly break the redshift-DM degeneracy and thereby substantially tighten the constraints on $\rm DM_{host}$ and $\sigma_{\rm host}$, as evident from the comparison between \autoref{fig:posa} and \autoref{fig:posd}. In addition, the baseband data with exact fluence measurements anchor the energy distribution and tighten the constraints on nearly all population parameters by removing the fluence uncertainty that otherwise broadens the inferred energy function. When the analysis is restricted to the CHIME/FRB Catalog~1 alone, our results are fully consistent with Ref. \cite{2023ApJShin}, who employed a logarithmically binned Poisson likelihood to characterize the FRB population. This agreement, obtained with an independent unbinned hierarchical Bayesian framework, provides a valuable cross check of our methodology and confirms that the inclusion of localized and baseband-detected bursts is the primary driver of the improved constraints presented in this work.

\begin{figure}
\centering
\includegraphics[width=0.95\linewidth]{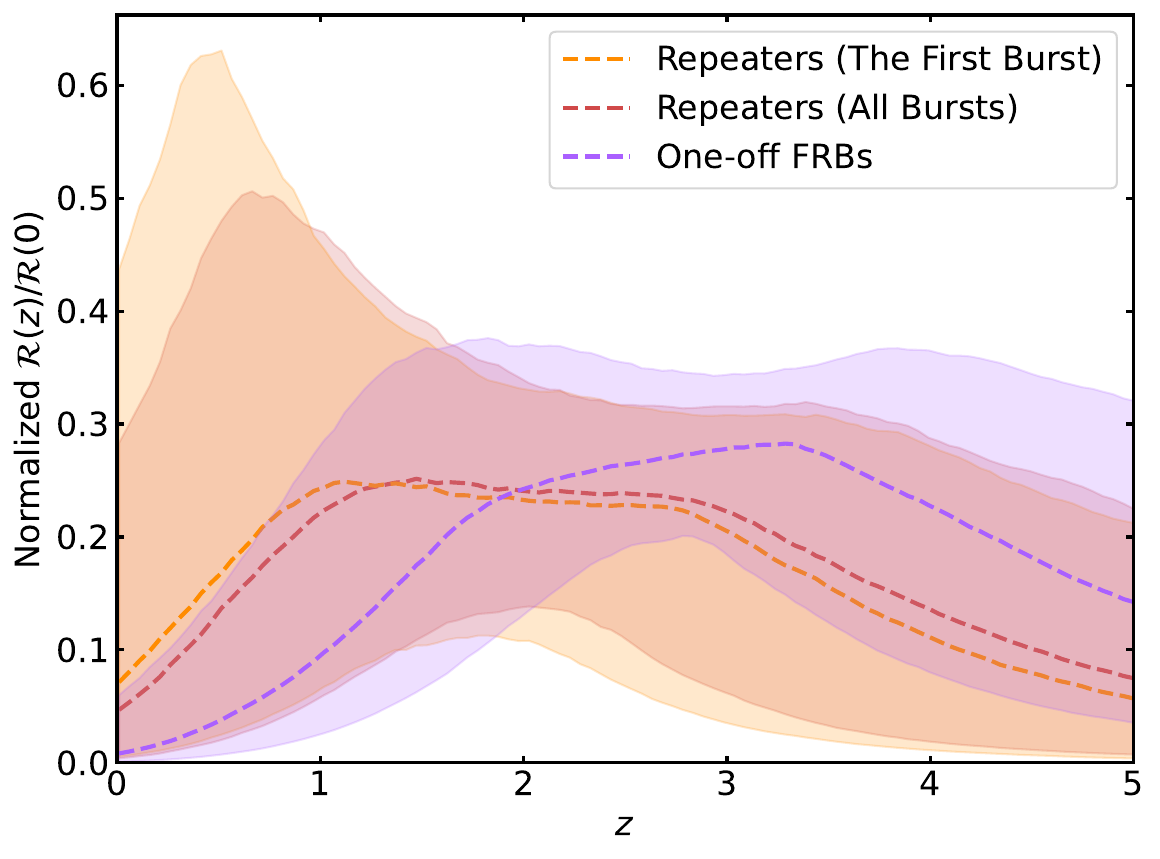}
\caption{Posterior distributions of the normalized rate $\mathcal{R}(z)/\mathcal{R}(0)$ under three treatments of repeating sources.}
\label{fig:rep_nonrep} 
\end{figure}

\begin{table}[!htbp]
\caption{\label{tab:repeaters}
Properties of the repeaters}
\begin{tabular}{lccc}
\hline
Names &S/N &$\rm DM_{\rm obs} \, (pc \, cm^{-3})$ &$F \, \rm (Jy \, ms)$\\
\hline
FRB20180917A &19.0 &189.92  &-\\
(FRB20180919A) &13.4 &188.85 &-\\
\hline
FRB20190103A &15.7 &364.56  &-\\
\hline
FRB20190208A &12.2 &579.84  &-\\
(FRB20190406A) &13.1 &580.23 &-\\
\hline
FRB20190222A &31.2 &460.17  &-\\
(FRB20190301A) &17.9 &459.44 &$13.5\pm1.5$\\
\hline
FRB20190417A &13.4 &1380.08 &-\\
\hline
FRB20190421A &12.3 &223.64  &-\\
\hline
FRB20190604A &33.8 &552.49  &-\\
\hline
FRB20190116A &15.8 &444.99  &$6.5\pm0.75$\\
(FRB20190116B) &12.9 &443.55 &-\\
\hline
FRB20190117A &24.2 &393.13 &$24\pm2.5$\\
(FRB20190630A) &19.4 &392.44 &-\\
\hline
\end{tabular}
\end{table}

\subsection{Repeaters and one-off sources}
For each repeating source, we retain the earliest burst that satisfies our selection criteria. Table~\ref{tab:repeaters} lists the properties we use. Three first detected bursts of the repeaters (FRB20180814A, FRB20181119A and FRB20190303A) are not considered because the S/N of their discovery burst is $<12$. None of the repeating sources has a confirmed redshift, and FRB20190116A, FRB20190117A, and the second burst of FRB20190222A (FRB20190301A) have fluence measurements. To verify that the repeaters do not affect our conclusions, we present the Madau-Extended results in \autoref{fig:rep_nonrep},  with (i) all  bursts from repeaters, (ii) the first burst of each repeater satisfying the criteria, and (iii) only one-off sources, respectively. Compared with \autoref{fig:prior_pos}, we find that the results with and without the repeaters are indistinguishable. Eight repeaters have $\rm DM_{\rm obs}<600 \, \rm pc \, cm^{-3}$, corresponding to $z\sim0.6$. Their posteriors are therefore not effectively constrained because of the small number of such sources. The parameter estimations of the one-off bursts are presented in \autoref{fig:posb}.

\begin{figure}
\centering
\includegraphics[width=0.95\linewidth]{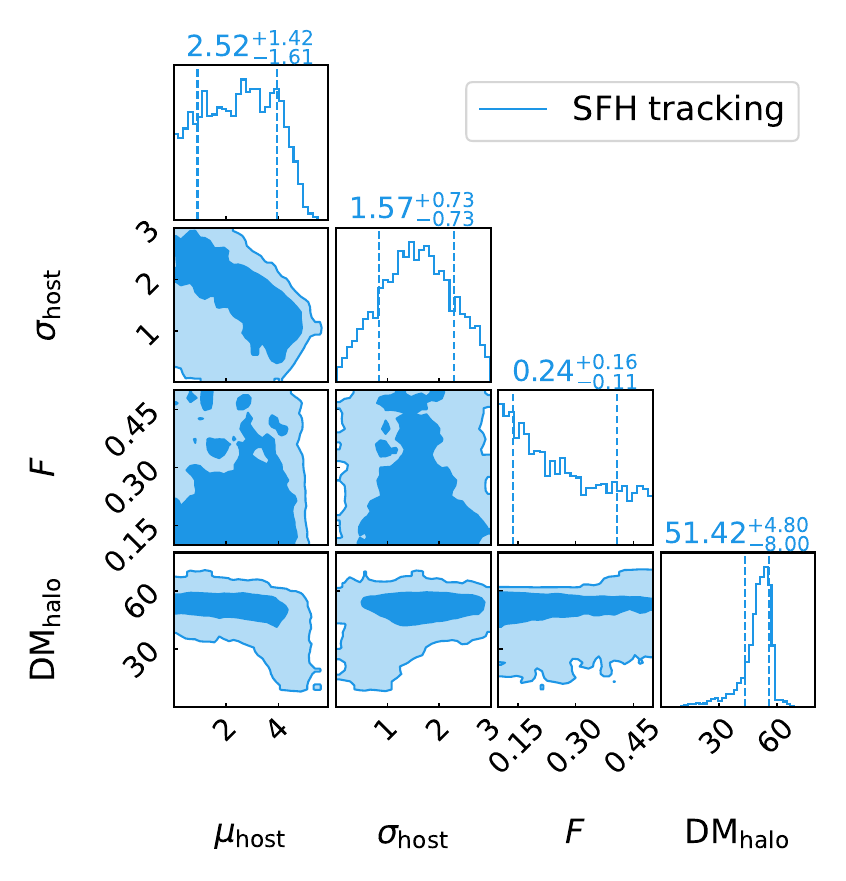}
\caption{Posterior distributions of the $\rm DM_{host}$, $\rm DM_{halo}$, and the feedback parameter $\mathcal{F}$ under the SFH tracking scenario.}
\label{fig:haloF} 
\end{figure}

\begin{table*}[!tp]
\caption{Sensitivity of the model parameters to the assumed Milky Way halo
contribution $\rm DM_{halo}$ and the feedback parameter $F$.}
\label{tab:halo_sensitivity}
\begin{tabular}{lccccc}
\hline
Parameter & $\rm DM_{halo} = 20$ & $\rm DM_{halo} =30$ (fiducial) & $\rm DM_{halo} =40$ & $\rm DM_{halo} =50$ & free $\rm DM_{halo}$ and $F$\\
\hline
$\log_{10}\Phi_0$ & $5.14^{+0.28}_{-0.25}$ & $5.11^{+0.28}_{-0.25}$ & $5.07^{+0.24}_{-0.23}$ & $5.05^{+0.20}_{-0.21}$ & $5.11^{+0.20}_{-0.22}$ \\
$\gamma$                 & $-1.87^{+0.14}_{-0.13}$ & $-1.85^{+0.14}_{-0.13}$ & $-1.83^{+0.13}_{-0.12}$ & $-1.81^{+0.11}_{-0.11}$ & $-1.83^{+0.12}_{-0.11}$ \\
$\log_{10}E_{\rm char}$  & $42.29^{+0.43}_{-0.46}$ & $42.26^{+0.43}_{-0.44}$ & $42.22^{+0.39}_{-0.43}$ & $42.19^{+0.38}_{-0.40}$ & $42.17^{+0.39}_{-0.42}$ \\
$n$                      & $1.57^{+0.64}_{-0.74}$ & $1.58^{+0.62}_{-0.71}$ & $1.56^{+0.62}_{-0.66}$ & $1.56^{+0.23}_{-0.28}$ & $1.41^{+0.66}_{-0.65}$ \\
$\alpha$                 & $-3.25^{+1.19}_{-1.11}$ & $-3.30^{+1.19}_{-1.07}$ & $-3.33^{+1.14}_{-1.06}$ & $-3.34^{+1.10}_{-1.02}$ & $-3.19^{+1.11}_{-1.09}$ \\
$\mu_{\rm host}$         & $4.34^{+0.51}_{-0.41}$ & $4.08^{+0.56}_{-0.58}$ & $3.60^{+0.74}_{-1.51}$ & $2.34^{+1.32}_{-1.53}$ & $2.52^{+1.41}_{-1.61}$ \\
$\sigma_{\rm host}$      & $0.67^{+0.41}_{-0.35}$ & $0.90^{+0.52}_{-0.47}$ & $1.22^{+0.74}_{-0.59}$ & $1.62^{+0.67}_{-0.68}$ & $1.57\pm0.73$ \\
$F$                      & $0.32$ & $0.32$ & $0.32$ & $0.32$ & $0.24^{+0.16}_{-0.11}$ \\
$\mathrm{DM_{halo}}$     & - & - & - & - & $51.42^{+4.78}_{-7.95}$ \\
\hline
\end{tabular}
\end{table*}

\subsection{The components of the observed dispersion measure}
Our decomposition $\mathrm{DM_{obs}}=\mathrm{DM_{ISM}}+\mathrm{DM_{halo}}+\mathrm{DM_{cosmic}}(z)+\mathrm{DM_{host}^{src}}/(1+z)$ separates each component of the observed DM. In the fiducial analysis, $\rm DM_{halo} =30\, pc \, cm^{-3}$ and the feedback parameter $\mathcal{F} =0.32$ in $\rm DM_{cosmic}$ are fixed. Both are degenerate with the host-DM parameters. Therefore, we repeat the analysis with $\mathrm{DM_{halo}}$ fixed at $20,\,30,\,40,\,50$~pc~cm$^{-3}$ ($F=0.32$), and with both $\mathrm{DM_{halo}}\sim\mathcal{U}(10,80)$~pc~cm$^{-3}$ and $F\sim\mathcal{U}(0.1,0.5)$ treated as free parameters. The results are summarized in Table~\ref{tab:halo_sensitivity} and \autoref{fig:haloF}. As expected, $\mu_{\rm host}$ decreases monotonically from $\sim4.3$ at $\mathrm{DM_{halo}}=20$~pc~cm$^{-3}$ to $\sim2.3$ at $\mathrm{DM_{halo}}=50$~pc~cm$^{-3}$, while $\sigma_{\rm host}$ increases from $\sim0.7$ to $\sim1.6$. The free run returns $\mathrm{DM_{halo}}\simeq52$~pc~cm$^{-3}$ and $\mathcal{F}\simeq0.24$. The feedback parameter is not tightly constrained. Other parameters that drive our conclusions are unchanged to within $\lesssim1\sigma$ across all configurations. 

After removing the Milky Way ISM contribution, the smallest remaining $\mathrm{DM}$ in the detected injection sample is $36.2$~pc~cm$^{-3}$. For a halo contribution exceeding this value, $\mathrm{DM_{cosmic}}+\mathrm{DM_{host}}$ would be negative for the lowest-$\mathrm{DM}$ events, so such events would be discarded. Given this limitation, we adopt $\mathrm{DM_{halo}}=30$~pc~cm$^{-3}$ as the fiducial value. To quantify the impact on our conclusion, we repeat the analysis with (i) free $\mathrm{DM_{halo}}$ and $F$, (ii) varied fixed $\mathrm{DM_{halo}}$, based on the fixed DTD model. We find that the $\Delta t_d$ constraints are stable for these scenarios, with $\Delta t_d \sim 0.09 \, \rm Gyr$, still favoring the assumed SFH. This indicates that our conclusion is not sensitive to the choice of $\mathrm{DM_{halo}}$ and $F$.

\section{End-to-end validation with mock catalogs}
\label{sec:mock_validation}
To further examine the reliability of our results, we perform end-to-end closure tests to verify that the analysis pipeline can recover the redshift evolution of an assumed FRB population, in particular for populations with long delay times. We use the catalog of detected injections as an empirical representation of the survey selection and recovery response. For each input population model, we fix the population parameters to ${\bf \Lambda}_{\rm inj}$ and adopt the same survey exposure and detection efficiency as in the analysis of the real data. In all scenarios we assume $\log_{10}\Phi_0 = 4.8$, $\gamma=-1.5$, $E_{\rm char}=10^{41.5}~{\rm erg}$, $\alpha=-1.5$, $\mu_{\rm host}=4.4$, and $\sigma_{\rm host}=0.5$. Since the Madau-Extended parametrization is the only model considered here that allows the intrinsic rate to peak above the SFH peak redshift, we test two scenarios, $(\lambda,\kappa,z_{\rm p})=(3.2,\,5.6,\,0.5)$ and $(2.7,\,5.6,\,3)$, representing a delayed population and a population peaking beyond the SFH, respectively.
For each realization, the size of the mock catalog is drawn from a Poisson distribution, $N_{\rm mock}\sim{\rm Poisson}(N_{\rm exp})$, consistent with the Poisson factor in the inhomogeneous-Poisson-process likelihood used in the inference. The $N_{\rm mock}$ detected injections are then resampled with replacement according to their model-dependent predictive weights,
\begin{equation}
p_j({\bf \Lambda}_{\rm inj})= \frac{w_j({\bf \Lambda}_{\rm inj})}{\sum_k w_k({\bf \Lambda}_{\rm inj})},
\end{equation}
where $w_j$ is given by $w_j \propto \pi(\theta_j|{\bf \Lambda})/P_{\rm inj}(\theta_j)$.

\begin{figure}
\centering
\includegraphics[width=0.95\linewidth]{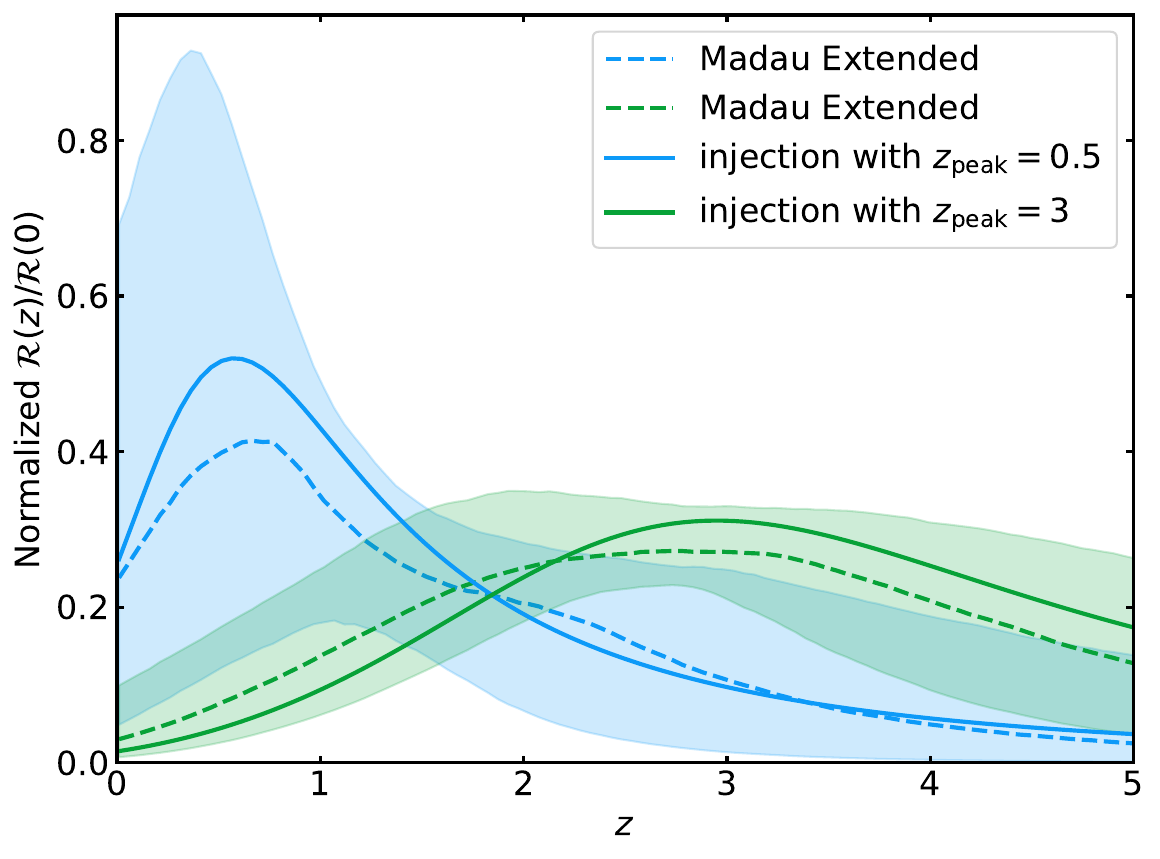}
\caption{End-to-end closure tests for the redshift evolution of the FRB volumetric rate using the flexible Madau-Extended parametrization. The solid blue and green curves show the injected normalized rate histories for mock populations with $z_{\rm peak}=0.5$ and $3$, respectively. The dashed curves show the posterior median reconstructed from the corresponding mock catalogs, and the shaded regions denote the $68\%$ credible intervals. }
\label{fig:mock_madau} 
\end{figure}

We also reproduce the observational availability pattern of the real sample. A fraction $4/225$ of the mock detections are assigned both a fluence and a redshift, an additional $79/225$ are assigned a fluence but no redshift, and the remaining events are assigned neither quantity. The localized events with known redshifts are drawn without replacement from the detections satisfying ${\rm DM}_{\rm inj}<1000~{\rm pc\,cm^{-3}}$ and ${F}_{\rm inj}>2~{\rm Jy\,ms}$. The events with fluence measurements are then drawn from the remaining detections, with a measurement uncertainty of $\sigma_F=0.1\,F$. For each localized event, the redshift is sampled from the conditional posterior
\begin{equation}
p(z|F_j,{\rm DM}_j,{\bf \Lambda}_{\rm inj}) \propto P(F_j|z,\Lambda_{\rm inj,1}) P({\rm DM}_j|z,\Lambda_{\rm inj,2}) p(z|\Lambda_{\rm inj,3}),
\end{equation}
so that the assigned redshifts are statistically consistent with the observed fluence and DM. The mock catalogs therefore reproduce both the model-dependent distribution of the detected observables and the availability pattern of the real observations. In all mock catalogs used here, the duplicate fraction is below $5\%$.

Both mock catalogs are then analyzed with exactly the same Bayesian likelihood and priors, and the same use of the fluence and redshift information, as in the analysis of the real data. \autoref{fig:mock_madau} shows the closure tests for the flexible Madau-Extended rate parametrization. The solid curves denote the input rate evolutions for mock populations with injected peaks at $z_{\rm peak}=0.5$ and $3$, whereas the dashed curves and shaded regions show the posterior reconstructions. In both cases, the recovered rate evolution follows the injected redshift dependence within the posterior uncertainties, demonstrating that the flexible parametrization is not intrinsically driven towards the canonical SFH peak. Consequently, the evidence against multi-Gyr delays inferred from the real catalog is not an artifact of the pipeline's inability to represent or recover such populations.

\renewcommand{\bibsection}{\par\vspace{\baselineskip}}

\bibliographystyle{apsrev4-1}
\bibliography{reference.bib}

\begin{figure*}
\includegraphics[width=0.9\linewidth]{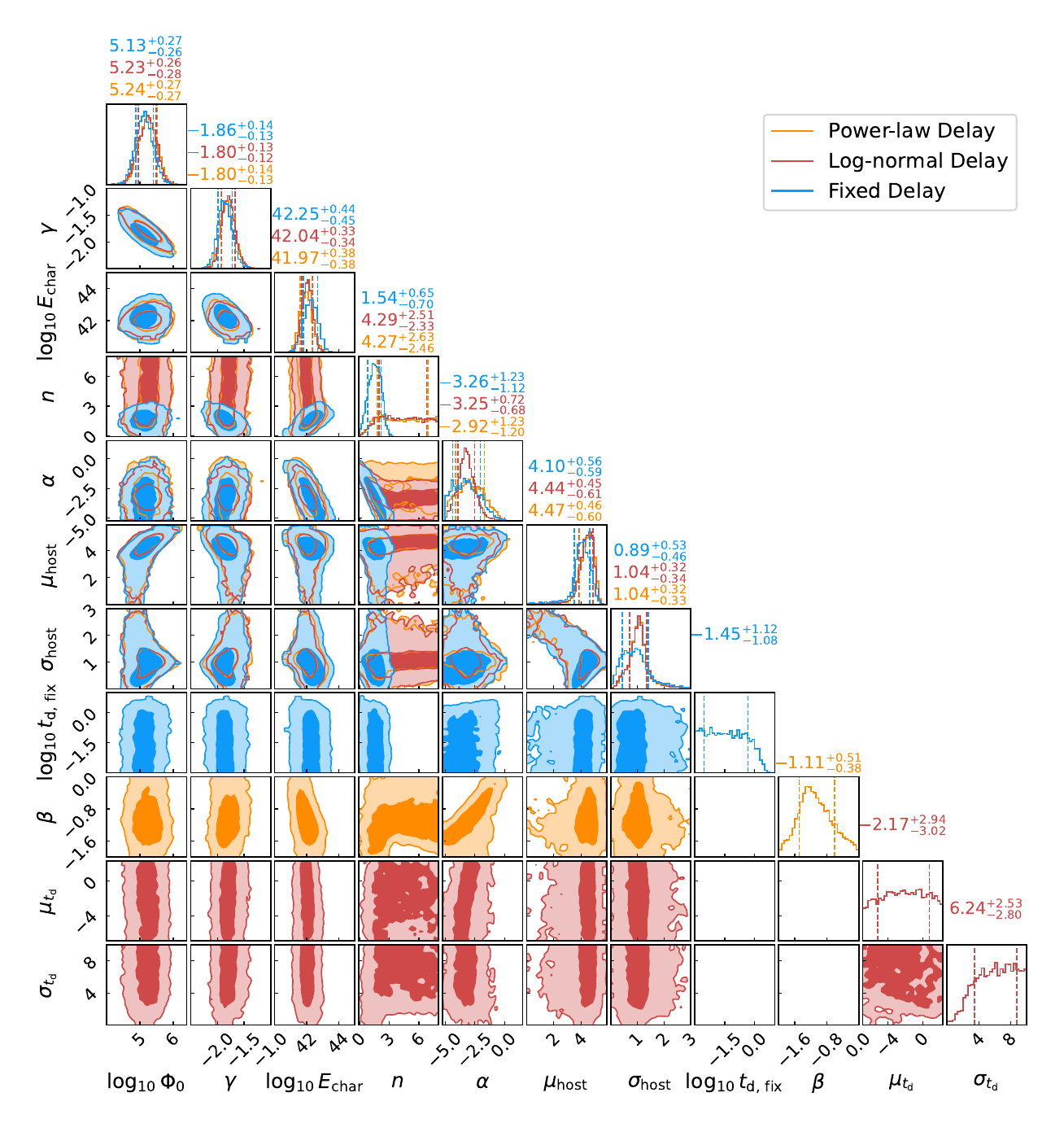}
\caption{Posterior distribution of hyper-parameters of the power-law delay, the log-normal delay, and the fixed delay models. The contours represent the 68\% and 99\% credible intervals, respectively. The values and dashed lines in the marginal distributions represent the 68\% credible intervals.}
\label{fig:posa} 
\end{figure*}

\begin{figure*}
\includegraphics[width=0.9\linewidth]{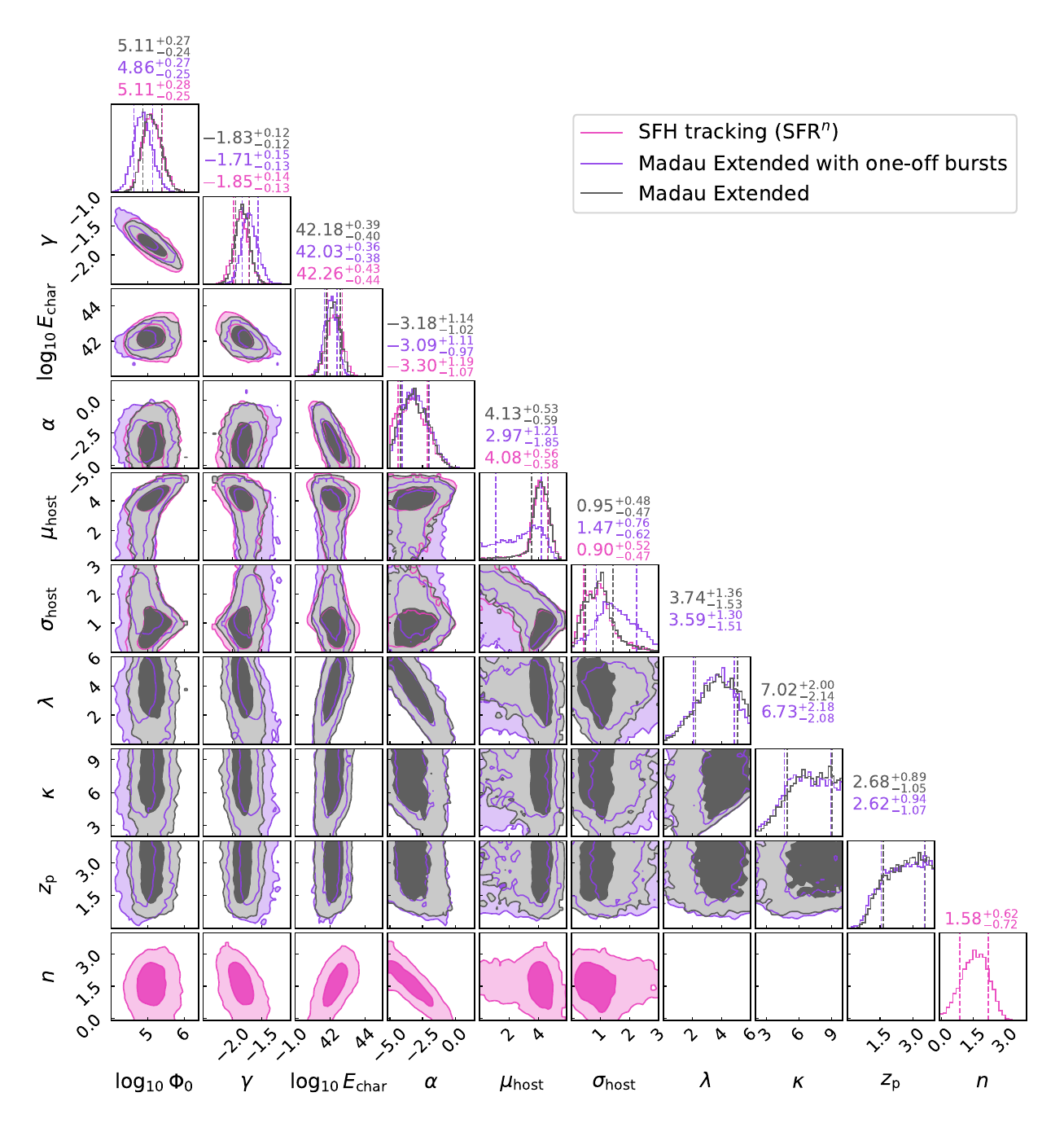}
\caption{Posterior distribution of hyper-parameters of the Madau-Extended model and the SFH tracking models using different data samples. The contours represent the 68\% and 99\% credible intervals, respectively. The values and dashed lines in the marginal distributions represent the 68\% credible intervals.}
\label{fig:posb} 
\end{figure*}

\begin{figure*}
\includegraphics[width=0.9\linewidth]{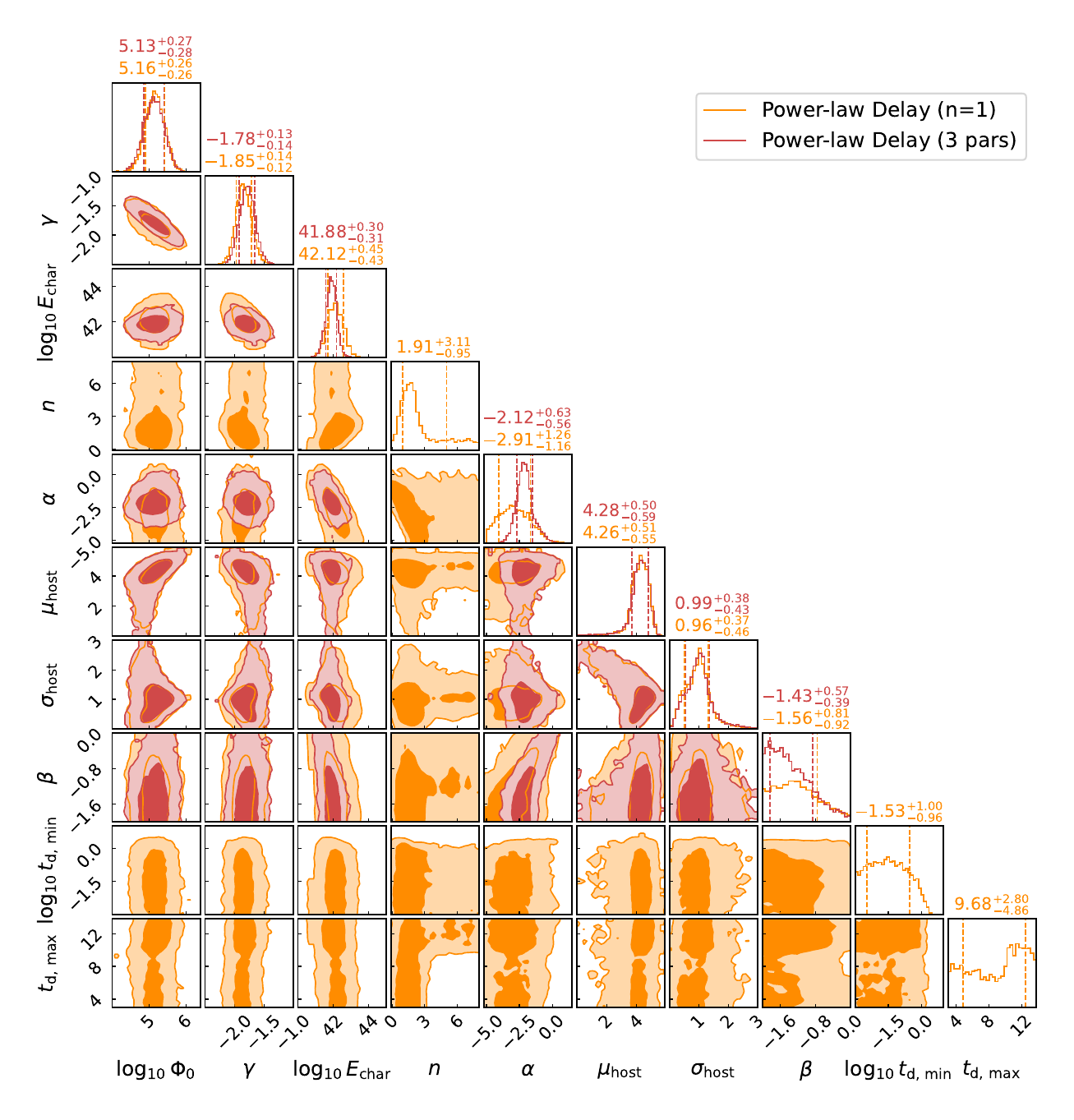}
\caption{Posterior distribution of hyper-parameters of the power-law delay model with different estimated parameters. The contours represent the 68\% and 99\% credible intervals, respectively. The values and dashed lines in the marginal distributions represent the 68\% credible intervals.}
\label{fig:posc} 
\end{figure*}

\begin{figure*}
\includegraphics[width=0.9\linewidth]{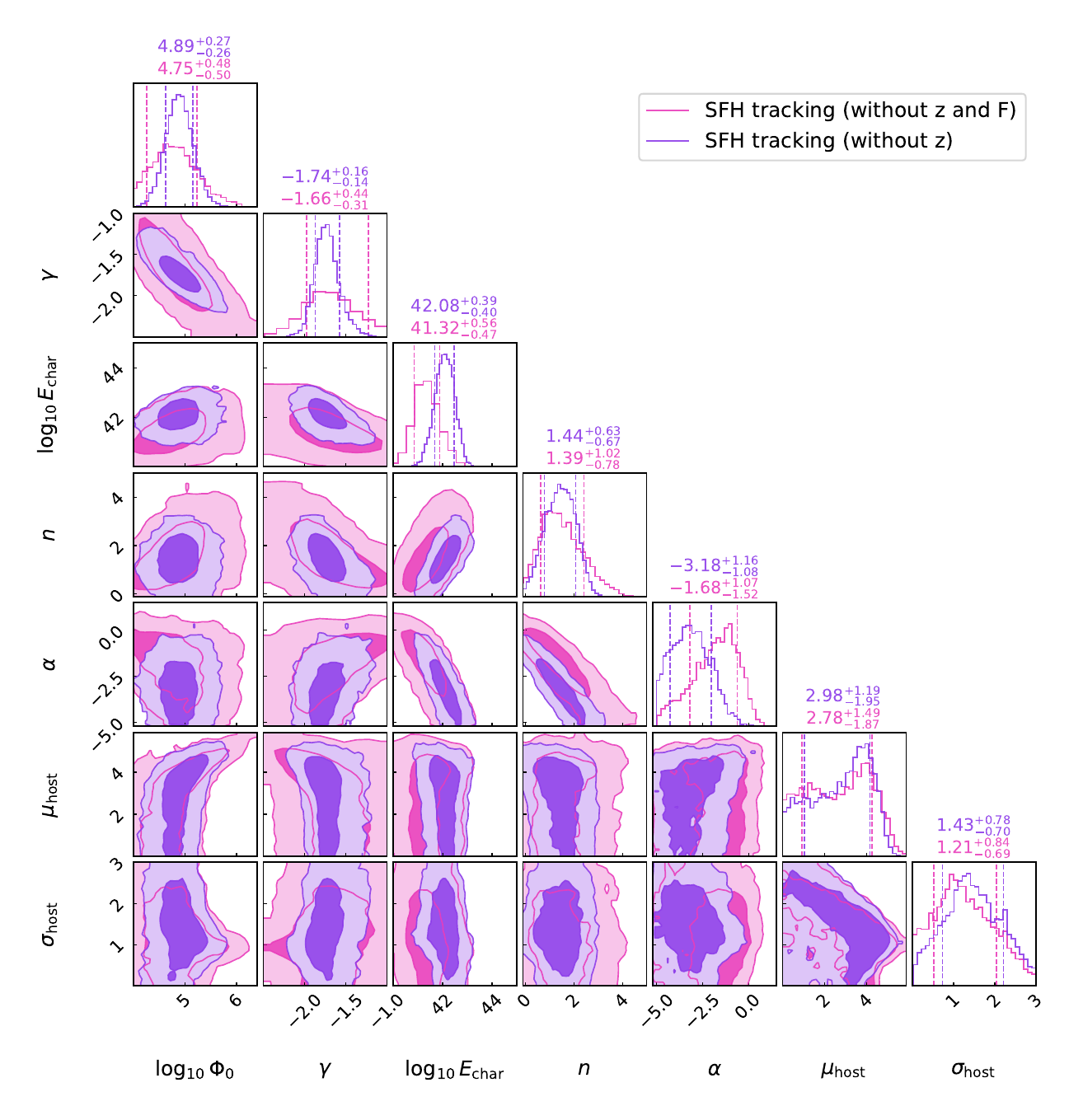}
\caption{Posterior distribution of hyper-parameters of the SFH tracking models using different types of the observations. The contours represent the 68\% and 99\% credible intervals, respectively. The values and dashed lines in the marginal distributions represent the 68\% credible intervals.}
\label{fig:posd} 
\end{figure*}